\documentclass[fleqn,usenatbib]{mnras}

\usepackage{newtxtext,newtxmath}

\usepackage[T1]{fontenc}

\usepackage{graphicx}	    %
\usepackage{amsmath}	    %
\usepackage{multicol}	    %
\usepackage{multirow}
\usepackage{bm} 	        %
\usepackage{pdflscape}	    %
\usepackage[normalem]{ulem} %
\usepackage{tensor, siunitx}
\usepackage{cuted}
\usepackage{comment}
\usepackage[dvipsnames]{xcolor}
\usepackage{subcaption}
\usepackage{enumitem}

\newcommand{\gtt}{g_{tt}}
\newcommand{\gtp}{g_{t\phi}}
\newcommand{\gpp}{g_{\phi\phi}}

\newcommand{\e}[1]{\times 10^{#1}}

\newcommand{\solarmass}{M_\odot}
\newcommand{\rhomax}{\rho_0^\mathrm{max}}
\newcommand{\ns}{_\mathrm{NS}}
\newcommand{\disk}{_\mathrm{Disk}}

\title[GR self-gravitating disks around NS]
{General relativistic self-gravitating equilibrium disks around rotating neutron
stars}

\author[Kim et al.]{
Yoonsoo Kim$^{1,2}$\thanks{\href{mailto:ykim7@caltech.edu}{ykim7@caltech.edu}},
Jinho Kim$^{3}$,
Hee Il Kim$^{4}$,
and Hyung Mok Lee$^{5}$\\
$^1$ Theoretical Astrophysics, Mailcode 350-17, California Institute of
Technology, Pasadena, CA 91125, USA \\
$^2$ Department of Physics \& Astronomy, Seoul National University, 1 Gwanak-ro,
Gwanak-gu, Seoul 08826, South Korea \\
$^3$ Korea Astronomy and Space Science Institute, 776 Daedeok-daero, Yuseong-gu,
Daejeon 34055, South Korea\\
$^4$ Center for Quantum Spacetime, Sogang University, 35 Baekbeom-ro, Mapo-gu,
Seoul 04107, South Korea\\
$^5$ Center for the Gravitational-Wave Universe, Astronomy Research Center,
Seoul National University, 1 Gwanak-ro, Gwanak-gu, Seoul 08826, South Korea
}

\date{Accepted XXX. Received YYY; in original form ZZZ}
\pubyear{2024}

\begin{document}
\label{firstpage}
\pagerange{\pageref{firstpage}--\pageref{lastpage}}
\maketitle

\begin{abstract}
    In modeling a relativistic disk around a compact object, the self-gravity of
    the disk is often neglected while it needs to be incorporated for more
    accurate descriptions in several circumstances. Extending the
    Komatsu-Eriguchi-Hachisu self-consistent field method, we present numerical
    models of a rapidly rotating neutron star with a self-gravitating disk in
    stationary equilibrium. In particular, our approach allows us to obtain
    numerical solutions involving a massive disk with the rest mass
    $\mathcal{O}(10^{-1})-\mathcal{O}(10^0)\,\solarmass$ closely attached to a
    rotating neutron star, given that the disk is mainly supported by the
    relativistic electron degeneracy pressure. We also assess the impact of
    self-gravity on the internal structure of the disk and the neutron star.
    These axisymmetric, stationary solutions can be employed for simulations
    involving the neutron star-disk system in the context of high-energy
    transients and gravitational wave emissions.
\end{abstract}


\section{Introduction}

Along with recent breakthroughs in multi-messenger astronomy
\citep{Abbott_2016,Abbott_2017,Kasliwal:2017ngb}, theoretical studies on compact
objects and the dynamics of relativistic matter surrounding them have a growing
significance. General relativistic approaches are required to properly describe
astrophysical phenomena taking place in a strong gravity regime, yet the high
complexity and nonlinearity of the Einstein field equations compel numerical
approaches to obtain solutions \citep{Baumgarte2010,Rezzolla2013}. Particularly
simple as well as useful are the equilibrium models of relativistic stars with
disks, which serve as not only a means to study long-term secular evolutions of
such systems but also as initial data of time-dependent numerical relativity
simulations. 

Numerical schemes for studying the structure of rotating neutron stars have been
developed in a variety of flavors (see e.g. \cite{Paschalidis2017} for a review
on this topic). Of particular interest is the work by \cite{Hachisu1986}, who
proposed a novel iterative self-consistent field method to construct equilibrium
models of rotating stars in the Newtonian framework. Based on a similar
strategy, \cite*{Komatsu1989a} constructed equilibrium models of uniformly
rotating neutron stars in full general relativity, and subsequently extended to
differential rotation \citep{Komatsu1989b}. This so-called the KEH method, which
is named after the three authors of the original work, has been widely employed
for studying rotating neutron stars
\cite[e.g.][]{Cook1992,Cook1994a,Cook1994b,Eriguchi1994,
Stergioulas1995,Nozawa1998,Bauswein:2017aur,Iosif2021,Iosif:2022} along with
some modifications made in later studies.

With the aim of modeling accretion disks around a black hole, the theory of an
equilibrium disk (torus) had been developed more in the context of a
non-self-gravitating fluid in a fixed background spacetime
\cite[e.g.][]{Fishbone1976,Kozlowski1978,Abramowicz1978} in its early stage. In
particular, disk solutions by \cite{Fishbone1976} have been extensively used to
initialize the matter distribution in black hole accretion simulations. In most
simulations of accreting compact objects, the self-gravity of the disk can be
often ignored to simplify the problem, owing to its small mass compared to the
central object \citep{Abramowicz:2011xu}.
However, this assumption might no longer be valid in some astrophysical
scenarios such as an accreting nascent compact object in core-collapse
supernovae \cite[e.g.][]{Fryer_1999,Chan:2017tdg,Burrows:2023nlq} or a massive
remnant disk formed after disruption of a neutron star during compact binary
merger events
\cite[e.g.][]{Foucart_2012,Foucart_2016,Radice:2018pdn,Bernuzzi:2020tgt,Kruger:2020gig,Camilletti2024}.

While early works on the relativistic, self-gravitating disks have mainly
explored black hole disks
\citep{Will:1974,Chakrabarti1988,Lanza1992,Nishida1994,Ansorg2005,Shibata2007},
it has been suggested that a massive disk can be formed around a neutron star
from the accretion-induced collapse of a white dwarf \citep{Abdikamalov:2009aq}
or binary neutron star mergers with asymmetric mass ratios \cite[$q\sim
0.75-0.85$, ][]{Rezzolla:2010fd}.

Equilibrium solutions of the neutron star-disk system can be used as initial
data for a range of astrophysical scenarios that can emerge from such a system.
Possible applications include investigating connections between X-ray
quasi-periodic oscillations and disk oscillation modes
\cite[e.g.][]{Blaes:2006fv,Fragile2018,matuszkova2024accretion-2}, hydrodynamic
instabilities of the self-gravitating disk \cite[see][for the black hole-disk
system]{Wessel:2020hvu}, the collapse of the disk via neutron star burst-disk
interaction \citep{Fragile2018}, disk-fed jet launching from a neutron star
\citep{Das:2023vow}, and accretion-induced collapse of the system into a black
hole \citep{Giacomazzo:2012bw}. Some of the aforementioned scenarios are
expected to launch electromagnetic signals such as gamma-ray bursts or emit
gravitational waves that can be observed with next-generation detectors.

Fully general relativistic models of a neutron star with a self-gravitating
disk, to the best of our knowledge, were first studied by \cite{Nishida1992}.
Based on the KEH scheme, they presented a set of numerical equilibrium solutions
of a uniformly rotating neutron star and a torus subject to the differential
rotation law suggested by \cite{Komatsu1989b}.
Albeit the self-gravity was neglected, we also mention the work of
\citet[\texttt{TORERO} code]{Corvino2009} which constructed equilibrium disk
around a rotating neutron star and was used to prepare initial data of
accretion-induced collapse simulations in \cite{Giacomazzo:2012bw}.

In this paper, we present equilibrium models of a rapidly rotating neutron star
with a self-gravitating disk based on the KEH scheme with an extension made to
construct a disk. Our formulation, specifically compared to the work of
\cite{Nishida1992}, uses the specific angular momentum as a basic variable
describing the rotation of the disk, providing a more direct and flexible
control over the angular momentum distribution of the disk. We also attempt to
quantitatively analyze the impact of the self-gravity of the disk on the neutron
star-disk equilibrium solutions.

This article is organized as follows. In Sec. \ref{sec:theory}, we review basic
equations describing general relativistic rotating fluid bodies in stationary
equilibrium. We introduce our iterative numerical scheme and its implementation
in Sec. \ref{sec:implementation}. Computed equilibrium models and discussions on
the disk self-gravity are presented in Sec. \ref{sec:results}, and we conclude
with a summary in Sec. \ref{sec:summary}. Throughout this paper we adopt
$c=G=\solarmass=1$ unit; conversion rules for some physical quantities are given
in Table \ref{tab:unit}. We use the abstract index notation with Latin indices
$(a, b, \cdots)$ for spacetime tensors.

\begin{table}
\caption{Units of some physical quantities when $c=G=\solarmass=1$.}
\label{tab:unit}
\begin{tabular}{lll}
    \hline
    Quantity & Conversion factor & Unit physical scale in cgs\\
    \hline\hline
    Length & $GM_\odot/c^2$ & $1.477\times 10^5\si{\cm}$ \\
    Time & $GM_\odot/c^3$ & $4.927 \times 10^{-6} \si{\second}$ \\
    Frequency & $c^3 / (GM_\odot)$ & $2.030\times 10^5\si{\hertz}$ \\
    Mass & $M_\odot$ & $1.988 \times 10^{33} \si{\gram}$ \\
    Mass Density & $c^6/(G^3 M_\odot^2)$ & $6.173 \times 10^{17}
    \si{\gram\per\cubic\cm}$ \\
    \hline
\end{tabular}
\end{table}

\section{Rotating fluid bodies in general relativity}
\label{sec:theory}

\subsection{Basic equations}

We consider a stationary and axisymmetric configuration of spacetime geometry
and matter distribution. The spacetime metric can be written as
\begin{equation}
\begin{split}\label{eq:metric}
    ds^2 & = \gtt dt^2 + g_{rr}dr^2 + g_{\theta\theta}d\theta^2
        + 2 \gtp dtd\phi + \gpp d\phi^2 \\[.5em]
         & = -e^{\gamma + \varrho}dt^2 + e^{2\alpha}(dr^2 + r^2 d\theta^2)
          + e^{\gamma - \varrho}r^2\sin^2\theta (d\phi - \omega dt)^2,
\end{split}
\end{equation}
where metric functions $\varrho, \gamma, \alpha, \omega$ depend on $r$ and
$\theta$ coordinates only. We model matter as the perfect fluid, which has the
stress-energy tensor
\begin{equation}\label{eq:perfect fluid stress-energy tensor}
    T^{ab} = (\rho_0 + \rho_i + p) u^a u^b + p g^{ab},
\end{equation}
where $\rho_0$ is rest energy density, $\rho_i$ is internal energy density, and
$p$ is pressure measured in the rest frame of the fluid. $u^a$ is the fluid
4-velocity. We assume the fluid follows a barotropic equation of state
$p=p\left(\rho_0\right)$. In this work, we only consider the polytropic law for
the equation of state, expressed as
\begin{equation} \label{eq:eos}
    p = K \rho_0^\Gamma
\end{equation}
where $K$ is the polytropic constant and $\Gamma$ is the polytropic exponent. It
follows from the first law of thermodynamics that
\begin{equation}
    \rho_i = \frac{p}{\Gamma - 1}.
\end{equation}
It is further assumed that the fluid is purely circularly rotating without any
meridional motion; the fluid 4-velocity has a form
\begin{equation} \label{eq:4-vel}
    u^a = (u^t, 0, 0, u^\phi) = u^t (1, 0, 0, \Omega),
\end{equation}
where $\Omega=u^\phi/u^t = d\phi/dt$ is the coordinate angular velocity. From
the normalization condition of the fluid 4-velocity ($u^a u_a = -1$), we get
\begin{equation}
    u^t
    = \big[- (\gtt + 2\gtp \Omega + \gpp \Omega^2)\big]^{-1/2}
    = \frac{e^{-(\gamma + \varrho)/2}}{\sqrt{1-V^2}},
\end{equation}
where $V = e^{-\varrho}(\Omega-\omega)r\sin\theta$.

From Eqs.~\eqref{eq:metric} and \eqref{eq:perfect fluid stress-energy tensor},
it is straightforward to evaluate the curvature tensor and write out the field
equations for each component. Following \cite{Komatsu1989a}, we use the
following form:
\begin{gather}
    \label{eq:keh-fieldeq-1}
    \nabla^2 ( \varrho e^{\gamma/2} ) = S_\varrho (g_{ab}, T_{ab}) \\[2mm]
    \label{eq:keh-fieldeq-2}
    \left[ \nabla^2 + \frac{1}{r}\frac{\partial}{\partial r} -
        \frac{1}{r^2}\mu\frac{\partial}{\partial \mu} \right]
        ( \gamma e^{\gamma/2} ) = S_\gamma (g_{ab}, T_{ab}) \\[2mm]
    \label{eq:keh-fieldeq-3}
    \left[ \nabla^2 + \frac{2}{r}\frac{\partial}{\partial r} -
        \frac{2}{r^2}\mu\frac{\partial}{\partial \mu} \right] ( \omega
    e^{(\gamma-2\varrho)/2} ) = S_\omega (g_{ab}, T_{ab}) \\[2mm]
    \label{eq:keh-fieldeq-4}
    \frac{\partial \alpha}{\partial \mu} = S_\alpha (g_{ab}, T_{ab})
\end{gather}
where $\mu = \cos \theta$ and $\nabla^2$ is the Laplacian operator in the flat
space. The detailed form of the source terms $S_\varrho, S_\gamma, S_\omega,
S_\alpha$ in Eq.~\eqref{eq:keh-fieldeq-1}-\eqref{eq:keh-fieldeq-4} can be looked
up from several references which include technical descriptions of the KEH
scheme \cite[e.g.][]{Komatsu1989a,Cook1992,Rezzolla2013}.

The equation of motion $\nabla_a \tensor{T}{^a^b} = 0$ gives the relativistic
Euler equation
\begin{equation}\label{eq:euler equation}
    - \partial_a \ln u^t + j(\Omega) \partial_a \Omega
    = - \frac{ \partial_a p}{\rho_0 + \rho_i + p} , 
\end{equation}
where
\begin{equation}
    j(\Omega) \equiv u^t u_\phi = \frac{l}{1-\Omega l}
\end{equation}
and
\begin{equation}
    l \equiv -\frac{u_\phi}{u_t} = - \frac{\gtp + \Omega\gpp}{\gtt + \Omega\gtp}
\end{equation}
is the specific angular momentum.

The relativistic von Zeipel's theorem \citep{Abramowicz1971} states that the
Euler equation \eqref{eq:euler equation} is integrable if equi-pressure and
equi-energy density surfaces coincide i.e., if the equation of state is
barotropic. Since the adopted polytropic equation of state \eqref{eq:eos}
satisfies the criterion, Eq.~\eqref{eq:euler equation} can be integrated into
\begin{equation}\label{eq:intEuler-basicform}
    -\ln u^t + \int j(\Omega)d\Omega + \ln H = C ,
\end{equation}
where $C$ is an integral constant and
\begin{equation} \label{eq:log enthalpy}
    \ln H \equiv \int \frac{dp}{\rho_0 + \rho_i + p}
    = \ln \left[ 1 + \frac{\Gamma}{\Gamma-1} K \rho_0^{\Gamma-1} \right].
\end{equation}

The relativistic Euler equation in its integral form
Eq.~\eqref{eq:intEuler-basicform} describes the matter distribution within a
single rotating fluid body, where the integration constant $C$ can be determined
by imposing proper boundary conditions. Depending on how the boundary conditions
are imposed, the fluid body (the region with $\ln H > 0$) can be in either a
spheroidal (star) or toroidal (disk or torus) configuration.

\subsection{Equilibrium solutions}

Specifying the functional form of $j(\Omega)$ amounts to choosing a specific
rotation profile of a fluid body. For rotating neutron star models, one of the
most widely used one is the one-parameter law $j(\Omega) = A^2 (\Omega_c -
\Omega)$ \citep{Komatsu1989a}, where $A$ is a free parameter that controls the
degree of differential rotation and $\Omega_c$ is angular velocity on the
rotation axis. In this work, we only consider the uniformly rotating neutron
star, for which Eq.~\eqref{eq:intEuler-basicform} simplifies to
\begin{equation} \label{eq:intEuler-star}
    \ln(H/u^t) = C .
\end{equation}

While the theory of a relativistic rotating star and an equilibrium torus have
all theoretical grounds in common apart from their topology (spheroidal or
toroidal), it is useful, when describing an equilibrium torus, to rewrite the
governing equations by changing the primary variable of rotation from angular
velocity $\Omega$ to the specific angular momentum $l$
\cite[e.g.][]{Kozlowski1978,Abramowicz1978,Font2002}. Using the following
relation
\begin{equation}\label{eq:l to omega}
    \Omega = -\frac{\gtp + \gtt l}{\gpp + \gtp l}
    \,,
\end{equation}
the relativistic Euler equation \eqref{eq:euler equation} can be rewritten to an
equivalent form
\begin{equation}\label{eq:Euler-anotherform}
    \partial_a \ln |u_t| + u_t u^\phi \partial_a l
    =  -\frac{\partial_a p}{\rho_0 + \rho_i + p}
    ,
\end{equation}
which can then be integrated to give
\begin{equation}\label{eq:intEuler-anotherform}
    \ln |u_t| + \int  u_t u^\phi d l + \ln H = C,
\end{equation}
where
\begin{equation}\label{eq:u_t}
    |u_t|^2 = \frac{\gtp^2 - \gtt\gpp}{\gtt l^2 + 2 \gtp l  + \gpp}
\end{equation}
and
\begin{equation}\label{eq:u_t u^phi}
    u_t u^\phi = - \frac{\Omega}{1-\Omega l}
    = \frac{\gtp + \gtt l}{\gtt l^2 + 2\gtp l + \gpp}.
\end{equation}

For a given background metric $g_{\mu\nu}$, three characteristic values of
specific angular momentum $l_\text{cr}$, $l_b$, $l_\text{K}$ can be defined as
follows. First, positivity of the denominator in Eq.~\eqref{eq:u_t} imposes the
following constraint on the specific angular momentum $l$:
\begin{equation} \label{eq:angular momentum critical bound}
    l_\text{cr}^-(r,\theta) < l(r,\theta) < l_\text{cr}^+(r,\theta)
\end{equation}
where
\begin{equation} \label{eq:angular momentum - critical}
    l_\text{cr}^\pm = \frac{\gtp \pm \sqrt{\gtp^2 - \gtt\gpp}}{-\gtt}.
\end{equation}
Matter is gravitationally bound to the system if $-u_t < 1$, which sets the
following condition
\begin{equation}
    l_\text{b}^-(r,\theta) < l < l_\text{b}^+(r,\theta)
\end{equation}
where
\begin{equation} \label{eq:angular momentum - bound}
    l_\text{b}^\pm = \frac{\gtp \pm \sqrt{(\gtp^2 - \gtt\gpp)(1+\gtt)}}{-\gtt}.
\end{equation}
Finally, on the equatorial plane, the Keplerian angular momentum is given as
\begin{equation} \label{eq:angular momentum - Kepler}
    l_\text{K}^\pm = \frac{-B \pm \sqrt{B^2-AC}}{A}
\end{equation}
where the coefficients $A$, $B$, and $C$ are given by
\begin{equation}
    \begin{split}
    A & = - \gtp^2 \frac{\partial \gtt}{\partial r}
        + 2 \gtt\gtp \frac{\partial \gtp}{\partial r}
        - \gtt^2 \frac{\partial \gpp}{\partial r}, \\
    B & = - \gtp\gpp \frac{\partial \gtt}{\partial r}
        + (\gtp^2 + \gtt\gpp) \frac{\partial \gtp}{\partial r}
        - \gtt\gtp\frac{\partial \gpp}{\partial r}, \\
    C & = -\gpp^2 \frac{\partial \gtt}{\partial r}
        + 2 \gtp\gpp \frac{\partial \gtp}{\partial r}
        - \gtp^2 \frac{\partial \gpp}{\partial r}.
    \end{split}
\end{equation}
If the matter is rotating with $l_\text{K}^\pm$ on the equatorial plane,
the motion is purely geodesic and the right hand side of Eq.~\eqref{eq:Euler-anotherform} vanishes (i.e. pressure gradient is zero).

\subsubsection{Von Zeipel's cylinders}
\label{sec:von zeipel cylinders}

\begin{figure}
\centering
\includegraphics[width=\linewidth]{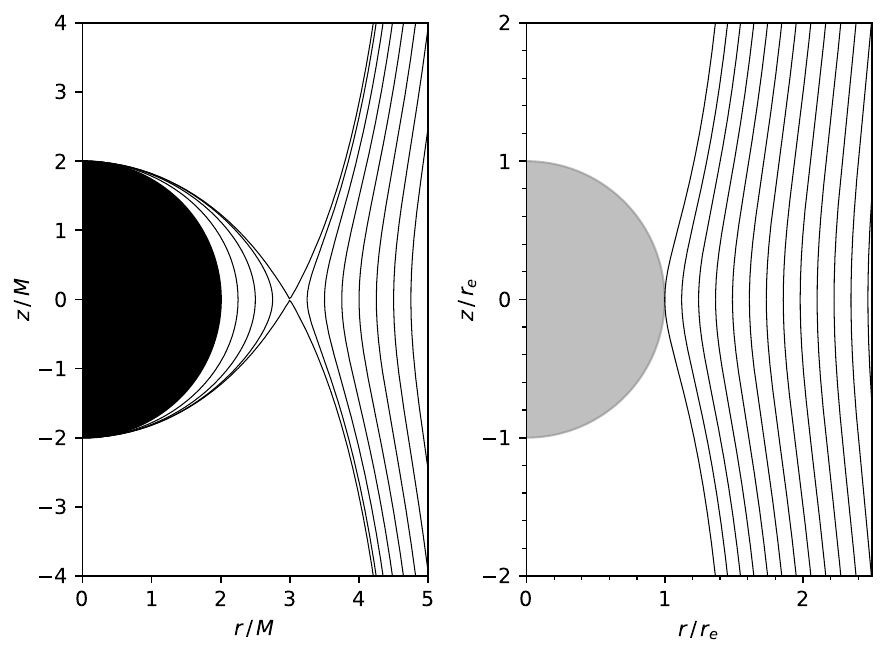}
\caption{The von Zeipel's cylinders (see Sec.~\ref{sec:von zeipel cylinders}) of a Schwarzschild black hole (left panel)
    and of a $M_0=1.8\solarmass$ spherical neutron star with the $K=100$,
    $\Gamma=2$ polytropic equation of state (right panel).}
\label{fig:zeipel cylinder}
\end{figure}

A further consequence of the relativistic von Zeipel's theorem
\citep{Abramowicz1971} is that when the Euler equation \eqref{eq:euler equation}
is integrable, const-$\Omega$ surfaces coincide with const-$l$ surfaces i.e.
$\Omega = \Omega(l)$. These surfaces are called the von Zeipel
cylinders\footnote{While these surfaces are cylinders in the Newtonian regime,
cylindrical topology is only asymptotic in general relativistic cases
\citep[][see also Figure~\ref{fig:zeipel cylinder}]{Abramowicz1974}.}. For a
specific value of $l_0$ and the corresponding value of $\Omega_0$ related by
Eq.~\eqref{eq:l to omega}, the equation
\begin{equation}\label{eq:Zeipel equation}
    \gtt (r,\theta)l_0 + (1 + \Omega_0 l_0) \gtp  (r,\theta) +	\gpp (r,\theta)
    \Omega_0 = 0 
\end{equation}
describes a single von Zeipel's cylinder characterized by $l_0$ (or
equivalently, by $\Omega_0$) in terms of the coordinates $(r,\theta)$.

When the specific angular momentum profile on the equatorial plane
$l_\text{eq}(r)$ is specified, Eq.~\eqref{eq:l to omega} gives the corresponding
profile of the angular velocity on the equatorial plane:
\begin{equation}\label{eq:omega along equatorial plane}
    \Omega_\text{eq}(r) = \Omega(r,\pi/2)
    = - \frac{\gtp(r,\pi/2) + \gtt(r,\pi/2) l_\text{eq}(r)}
    {\gpp(r,\pi/2) + \gtp(r,\pi/2) l_\text{eq}(r)} .
\end{equation}
Combining Eq.~\eqref{eq:Zeipel equation} and \eqref{eq:omega along equatorial
plane}, a von Zeipel's cylinder intersecting the equatorial plane at the radius
$r_0$ is described by an equation
\begin{equation}\label{eq:Zeipel equation - expanded}
    \begin{split}
        & l^2_\text{eq}(r_0) [ \gtp(r_0,\pi/2)\gtt(r,\theta) -
            \gtt(r_0,\pi/2)\gtp(r,\theta) ]
        \\ & \quad + l_\text{eq}(r_0) [ \gpp(r_0,\pi/2)\gtt(r,\theta) -
            \gtt(r_0,\pi/2)\gpp(r,\theta) ]
        \\ & \quad + [ \gpp(r_0,\pi/2)\gtp(r,\theta) -
        \gtp(r_0,\pi/2)\gpp(r,\theta) ] = 0.
    \end{split}
\end{equation}
For example, in the Schwarzschild spacetime, Eq.~\eqref{eq:Zeipel equation -
expanded} simplifies to
\begin{equation}
    (r_0-2M)r^3\sin^2\theta - r_0^3(r-2M) = 0.
\end{equation}
See e.g. \cite{Daigne2004} for the structure of von Zeipel's cylinders in the
Kerr spacetime for different values of the black hole spin. When an analytic
expression of the spacetime metric is not available, Eq.~\eqref{eq:Zeipel
equation - expanded} needs to be numerically solved to obtain the structure of
the von Zeipel's cylinders.

In Figure \ref{fig:zeipel cylinder}, we show the von Zeipel's cylinders outside
the horizon of a Schwarzschild black hole and the surface of a spherical neutron
star with the rest mass $M_0=1.8\solarmass$ and $K = 100$, $\Gamma=2$ polytropic
equation of state. Note that the cusp at $r=3M$ for the black hole (left panel
of Figure~\ref{fig:zeipel cylinder}) does not appear for the neutron star (right
panel).

\subsubsection{Equipotential surfaces and matter distribution}
\label{sec:equipotential theory}

The relativistic Euler equation in an alternative form
\eqref{eq:intEuler-anotherform} can be rearranged as
\begin{equation}
    W = C - \ln H,
\end{equation}
where the effective potential $W$ is defined as
\begin{equation}\label{eq:effective potential}
    W = \ln|u_t| - \int \frac{\Omega \, dl}{1-\Omega l} .
\end{equation}
Applying Eq.~\eqref{eq:effective potential} between two points $(r, \pi/2)$ and
$(\infty, \pi/2)$ on the equatorial plane,
\begin{equation} \label{eq:effective potential - equatorial plane}
    \begin{split}
        W_\text{eq}(r) & = W(r,\pi/2)
        \\ & = \ln |u_t(r, \pi/2)| - \int_\infty^{r}
        \frac{\Omega_\mathrm{eq}}{1-\Omega_\mathrm{eq} l_\mathrm{eq}}
        \frac{dl_\mathrm{eq}(r)}{dr}dr
    \end{split}
\end{equation}
where we use a fiducial offset $W=0$ at the infinity. If there is a von Zeipel's
cylinder passing through two points $(r,\theta)$ and $(r_0, \pi/2)$, since it is
a constant-$l$ surface, using Eq.~\eqref{eq:effective potential} and
\eqref{eq:effective potential - equatorial plane} we get
\begin{equation}
\begin{split}
    W(r,\theta) & = W_\text{eq}(r_0) + \ln \frac{u_t(r,\theta)}{u_t(r_0,\pi/2)}
        \\[.5em]
     & = \ln |u_t(r,\theta)| - \int_\infty^{r_0}
     \frac{\Omega_\mathrm{eq}}{1-\Omega_\mathrm{eq} l_\mathrm{eq}}
     \frac{dl_\mathrm{eq}(r)}{dr}dr .
\end{split}
\end{equation}
Note that the variable $r_0$ in the equation \eqref{eq:effective potential -
equatorial plane} is a function of $(r,\theta)$, which depends on the detailed
geometry of the von Zeipel's cylinders in a given spacetime. Plugging in
Eq.~\eqref{eq:u_t} and \eqref{eq:u_t u^phi}, we have
\begin{equation} \label{eq:effective potential - computing form}
\begin{split}
    W(r,\theta) &= \frac{1}{2}\ln\frac{\gtp^2 - \gtt\gpp}{\gtt l^2+2\gtp l+\gpp}
    \\[.5em] & \hspace{2em} + \int_\infty^{r_0(r,\theta)}
    \frac{\gtp + \gtt l_\text{eq}}{\gtt l_\text{eq}^2 + 2\gtp l_\text{eq} + \gpp}
    \frac{dl_\mathrm{eq}(r)}{dr}dr
\end{split}
\end{equation}

Given a spacetime metric and an angular momentum profile $l_\text{eq}(r)$ on the
equatorial plane, one can obtain the matter distribution of an equilibrium disk
through the following procedure. 1) Solve the equation \eqref{eq:Zeipel equation
- expanded} at each points to find $r_0(r,\theta)$, which is the coordinate
radius of the point at which the von Zeipel's cylinder passing through a point
$(r,\theta)$ intersects the equatorial plane. 2) Determine $l(r,\theta) =
l_\text{eq}(r_0)$, so one can compute $W(r,\theta)$ with Eq.~\eqref{eq:effective
potential - computing form}. 3) Choose the radius of the inner edge of the disk
$r=r_\text{in}$ on the equatorial plane to determine $W_\text{in} =
W(r_\text{in}, \pi/2)$, which amounts to imposing the boundary condition of the
Euler equation \eqref{eq:intEuler-anotherform}. 4) Compute the specific enthalpy
at every point using
\begin{equation} \label{eq:log H disk from effective potential}
    \ln H (r,\theta) = W_\text{in} - W(r,\theta)
\end{equation}
to recover the rest energy density from Eq.~\eqref{eq:log enthalpy}.

In this work, we assume that the distribution of the specific angular momentum
on the equatorial plane follows the power law \cite[e.g.][]{Daigne2004}
\begin{equation}\label{eq:disk rotation law}
    l_\mathrm{eq}(r) \equiv l(r, \pi/2) = \kappa \left(\frac{r}{r_e}\right)^a
\end{equation}
within the disk, where $\kappa$ is a constant, $a \geq 0$ is a power index, and
$r_e$ is the coordinate radius of the neutron star.

\section{Numerical implementation}
\label{sec:implementation}

\begin{table}
\centering
\caption{List of parameters specifying a single equilibrium model.}
\label{tab:model parameters}
\begin{tabular}{ p{0.2\linewidth} | p{0.65\linewidth} }
\hline
    Parameters & Description \\
\hline
    $\rhomax$ & Neutron star maximum rest mass density \\
    $r_p/r_e$ & Neutron star axis ratio \\
    $\kappa, a$ & Angular momentum distribution of the disk on the
        equatorial plane \\
    $W_\text{in}$ (or $r_\text{in}$) & Inner edge of the disk \\
\hline
\end{tabular}
\end{table}

\subsection{Computational procedure}
\label{sec:method outline}

In this section, we describe our iterative method for constructing a numerical
equilibrium model of a neutron star with a self-gravitating disk. Hereafter
subscripts "NS" and "Disk" are used to denote quantities related to the neutron
star and the equilibrium disk, respectively.

First, a numerical model of an isolated rotating neutron star is constructed
following the KEH scheme \cite[e.g.][]{Komatsu1989a,Cook1992}. This requires two
parameters fixed: the maximum rest energy density $\rhomax$ and the neutron star
axis ratio $r_p/r_e$. Once we converge to a rotating neutron star model with the
desired values of $(\rhomax, r_p/r_e)$ in the computational space, we choose the
rotation parameters $(\kappa, a)$ to fix the specific angular momentum
distribution $l_\text{eq}(r)$ of the disk on the equatorial plane (see
Sec.~\ref{sec:bounds on disk parameters} for the details). Keeping all the
parameters $(\rhomax, r_p/r_e, \kappa, a)$ fixed, each iteration step proceeds
as follows:

\begin{enumerate}[leftmargin=2em, labelwidth=1em, itemsep=0.5ex]
\item Make an initial guess on spacetime metric and matter distribution, which
    we denote by $g^{(n)}$ and $T^{(n)}$, where $n$ is the iteration number.
\item Evaluate the source terms in field equations \eqref{eq:keh-fieldeq-1} -
    \eqref{eq:keh-fieldeq-4} using $g^{(n)}$ and $T^{(n)}$, then integrate the
    field equations to compute updated metric $g^{(n+1)}$.
\item \emph{Building a neutron star} : Apply hydrodynamic boundary conditions
    for the neutron star by applying Eq.~\eqref{eq:intEuler-star} simultaneously
    at the pole, the equator, and the center of the neutron star. The coordinate
    radius $r_e^{(n+1)}$ and the angular velocity $\Omega_0^{(n+1)}$ of the
    neutron star are determined in this step. Then compute $\ln
    H_\text{NS}(r,\theta)$ with the equation \eqref{eq:intEuler-star} to obtain
    an updated matter distribution $T_\text{NS}^{(n+1)}$ of the neutron star.
\item \emph{Building a disk} : Follow the procedure outlined in
    Sec.~\ref{sec:equipotential theory} to construct an equilibrium disk. At the
    grid points outside the surface of the neutron star, solve
    Eq.~\eqref{eq:Zeipel equation - expanded} to get the values of
    $r_0(r,\theta)$\footnote{While $l_\text{eq}(r)$ is given as a continuous
    function by Eq.~\eqref{eq:disk rotation law}, values of $\gtt$, $\gtp$, and
    $\gpp$ are given discretely on the grid points on the equatorial plane. We
    employ cubic Hermite spline to interpolate metric coefficients in the root
    finding.} and compute the effective potential $W(r,\theta)$ using
    Eq.~\eqref{eq:effective potential - computing form}. Pick a value of
    $W_\mathrm{in}$, (or equivalently, pick the inner edge radius $r_\text{in}$
    of the disk at which $W_\text{in} \equiv W_\text{eq}(r_\text{in})$), then
    compute $\ln H\disk$ with Eq. \eqref{eq:log H disk from effective potential}
    to get the updated matter distribution $T\disk^{(n+1)}$ of the disk.
\item Feed $g^{(n+1)}$ and $T^{(n+1)} = T^{(n+1)}\ns + T^{(n+1)}\disk$ into the
    step (i) as the initial guess for the next iteration.
\end{enumerate}
In the step (iii) and (iv), we set $\rho_0=0$ in the regions at which $\ln
H(r,\theta) \leq 0$.

Steps (i)-(v) are repeated until numerical values of physical quantities
converge. In particular, we use the following convergence criteria
\begin{equation}\label{eq:convergence criteria}
    \Bigg | \frac{ \mathrm{max}(Q^{(n+1)} - Q^{(n)}) }
    {\mathrm{max}( Q^{(n)} )} \Bigg | < \Delta
\end{equation}
with $\Delta = 10^{-9}$ for a physical quantity $Q(r,\theta)$. The criteria
\eqref{eq:convergence criteria} is applied to the metric functions $\varrho,
\gamma, \omega, \alpha$ and the rest energy density $\rho_0$. Computation is
terminated if code fails or the number of iterations exceeds $N_\text{max} =
2000$.

Before entering the cycle of iterations (i)-(v) for the first time, we start
with a value of $W_\text{in}$ very close to the minimum value of
$W_\text{eq}(r)$ to converge to a solution with a disk of a very small mass
first. Then the value of $W_\text{in} - W_\text{center}$ is increased by a small
increment to perform another set of iterations until the solution converges to a
new state with a slightly larger disk mass. In other words, we start with a
light disk and use it as an initial guess (seed solution) for a heavier disk
progressively. By repeating this procedure, a sequence of numerical models with
increasing disk mass can be obtained.
Our strategy outlined above requires multiple rounds of iterations to reach a
model with a massive disk, starting from and iterating through the ones with a
lighter disk. While directly approaching to a single point chosen in the
parameter space can be seemingly computationally more efficient, we have
empirically observed that if the resulting disk is too massive, iterations
become quickly unstable and fail within few cycles.

Table~\ref{tab:model parameters} summarizes the set of parameters specifying a
single equilibrium solution in our scheme.

\subsection{Bounds on disk rotation parameters}
\label{sec:bounds on disk parameters}

\begin{figure}
{\includegraphics[width=\linewidth]{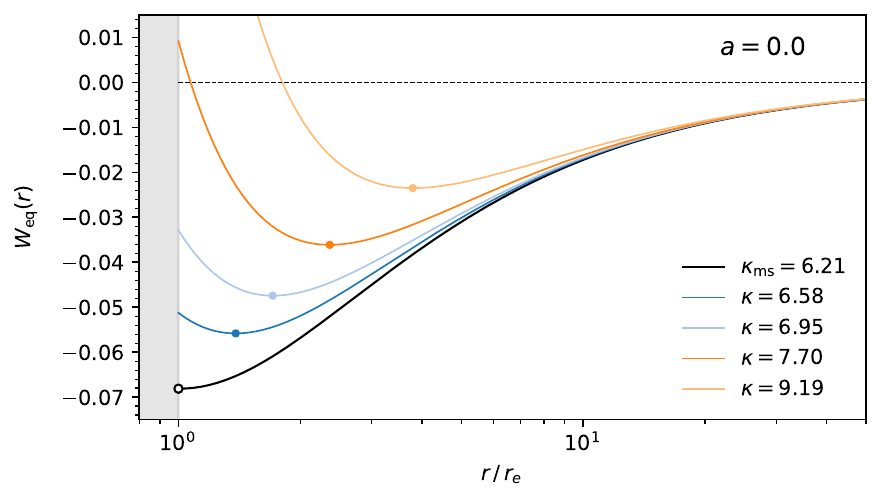}}
{\includegraphics[width=\linewidth]{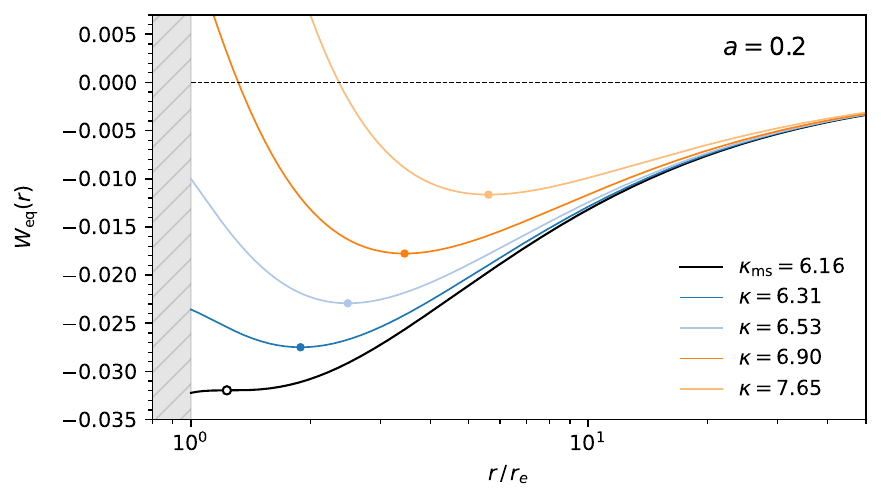}}
{\includegraphics[width=\linewidth]{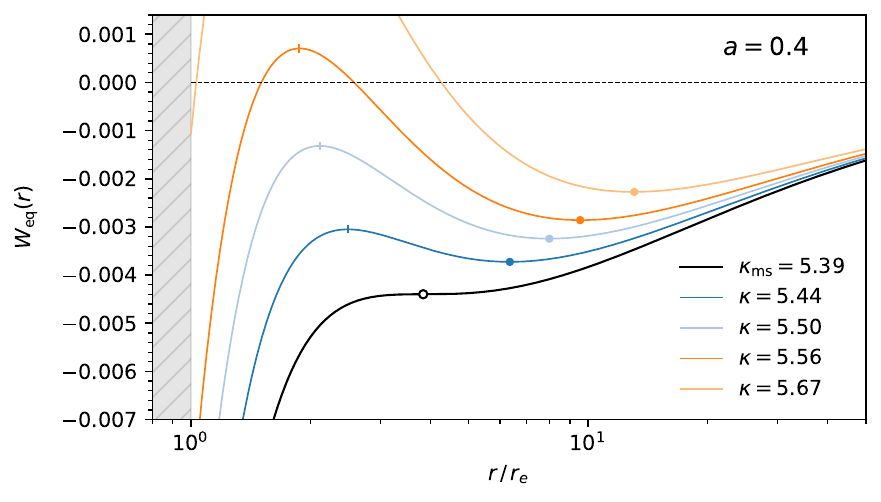}}
\caption{Variation of the effective potential $W_\text{eq}(r)$ with the disk
    rotation constant $\kappa$ and the power index $a$. The background spacetime
    is a rotating neutron star model with the maximum rest energy density
    $\rhomax=2.0\e{-3}$ and the axis ratio $r_p/r_e = 0.60$. In each panel, we
    show all types of stationary points $dW_\text{eq}/dr=0$ on curves: center
    with a filled dot ($\bullet$), inflection point with an empty dot ($\circ$),
    and cusp with a short vertical line ($\vert$) if present.}
\label{fig:effective potential without disk}
\end{figure}

Specification of the rotation law in Eq. \eqref{eq:disk rotation law} with a
boundary condition uniquely determines a configuration of the disk. In this
section, we describe our strategy for choosing a viable set of disk rotation
parameters $(\kappa, a)$ and $W_\text{in}$.

It is expected that, as the mass of the disk becomes comparable with the neutron
star, the disk begins to manifest its self-gravity. An equilibrium solution will
start to nonlinearly deviate from a simple superposition of a rotating neutron
star model with a non-self-gravitating disk. On the other hand, in the $M\disk
\ll M\ns$ limit, the equilibrium solution approaches a non-self-gravitating
added on top of a rotating neutron star solution, with a negligible contribution
to the overall spacetime geometry. The latter is a regime where the analytic
theory of a non-self-gravitating disk \cite[e.g.][]{Fishbone1976,Abramowicz1978}
is applicable, neglecting the contribution of $T\disk^{\mu\nu}$ in field
equations.

As described in Sec.~\ref{sec:method outline}, our method of constructing a
star-disk model begins with preparing an equilibrium solution of an isolated
rotating neutron star and superposing a very small disk on top of it. This small
disk is located at the local minimum (center) of the effective potential
$W_\text{eq}(r)$. Before we proceed with performing iterations to find an
equilibrium solution for star-disk, investigating the behavior of the effective
potential $W_\text{eq}(r)$ with different disk rotation parameters $\kappa$ and
$a$ can provide insights into a generic trend in self-gravitating solutions.

For a disk with a closed surface to exist, the effective potential $W(r,\theta)$
needs to have a local minimum outside $r=r_e$. In Figure \ref{fig:effective
potential without disk}, we show how the effective potential $W_\text{eq}(r)$
varies with the disk rotation constant $\kappa$ and the power index $a$. The
background spacetime is an isolated rotating neutron star model with the maximum
rest energy density $\rhomax=2.0\e{-3}$ and the axis ratio $r_p/r_e=0.60$.
Following the classification in \cite{Font2002}, we identify three types of
stationary points at which $dW_\text{eq}/dr = 0$:
\begin{itemize}[leftmargin=2em, labelwidth=1em]
    \item Center ($d^2W_\text{eq}/dr^2 > 0$, local minimum),
    \item Cusp ($d^2W_\text{eq}/dr^2 < 0$, local maximum),
    \item Inflection point ($d^2W_\text{eq}/dr^2 = 0$).
\end{itemize}
These points are also displayed in Figure~\ref{fig:effective potential without
disk} with different marker symbols. Note that $l_\text{eq} = l_\text{K}$ at
these stationary points.

Some qualitative observations can be made from Figure~\ref{fig:effective
potential without disk}. First, the cusp and inflection points are not present
for constant-angular momentum disks ($a=0.0$) and begin to appear at a higher
value of $a$. Next, the effective potential $W_\text{eq}(r)$ becomes more flat
and shallow with higher $a$. Note that the Keplerian specific angular momentum
$l_\text{K}$ asymptotically approaches to $a=0.5$ as $r\rightarrow\infty$, at
which $W_\text{eq}(r)$ converges to a horizontal flat curve. Finally, the
overall shape of $W_\text{eq}(r)$ changes more sensitively with respect to
$\kappa$ for higher $a$. These trends can be compared with a black hole
spacetime in which the cusp is present even with constant angular momentum
($a=0$) disk \cite[e.g. see][]{Font2002}.

If $\kappa$ is too small, the effective potential does not feature a center and
it is not possible to construct a small disk as an initial guess (recall that
our approach begins to construct a disk with a very small mass and moves to
progressively larger masses). In contrast, if $\kappa$ is too large, the disk
will be located very far away from the neutron star and the model may be of
little astrophysical interest. To this end, we choose to consider $\kappa$
within the interval
\begin{equation} \label{eq:kappa range}
    \kappa \in [\kappa_\text{ms}, \kappa_c].
\end{equation}

The lower bound $\kappa_\text{ms}$\footnote{The subscript stands for `marginally
stable' \citep{Font2002}.} is found by investigating the range of $\kappa$ for
which the effective potential $W_\text{eq}(r)$ has a center within $r_e < r <
\infty$. If an inflection point exists for some value of $\kappa$ (middle and
lower panel of Figure~\ref{fig:effective potential without disk}), we adopt it
as $\kappa_\text{ms}$. If an inflection point does not exist, $\kappa_\text{ms}$
equals the value of $\kappa$ at which the center is located at
$r=r_e$\footnote{The center is a stable point ($d^2W_\text{eq}/dr^2 > 0$) in
this case, but we will use the same symbol $\kappa_\text{ms}$ for it.} (top
panel of Figure~\ref{fig:effective potential without disk}). In practice,
$\kappa_\text{ms}$ can be determined by picking an initial value of $\kappa$
possessing a center in the corresponding $W_\text{eq}(r)$ and gradually
decreasing $\kappa$ until no center is found in $r > r_e$.

The upper bound $\kappa_c$ in \eqref{eq:kappa range} is simply set with the
condition $l_\text{cr}(r_e) = l_\text{eq}(r_e)$:
\begin{equation}\label{eq:kappa_critical}
    \kappa_c = \left(\frac{\gtp + \sqrt{\gtp^2 - \gtt\gpp}}{-\gtt}
    \right)_{r=r_e, \theta=\pi/2}.
\end{equation}
Empirically we find that Eq.~\eqref{eq:kappa_critical} serves as a reasonable
upper cutoff for the parameter space of our interest, and provides a practical
estimate on how much the constant $\kappa$ needs to be varied to modify the
configuration of the disk.

\subsection{Parametrization of disk mass}
\label{sec:parametrizing disk mass}

In the step (iv) described in Sec.~\ref{sec:method outline}, the location of the
inner edge of the disk needs to be specified during iterations either in terms
of $W_\text{in}$ or $r_\text{in}$. This amounts to controlling the maximum rest
energy density of the disk, which is determined from
\begin{equation}
    W_\text{in} - W_\text{center} = \ln [\max(H\disk)] .
\end{equation}
For the iterative procedure to stably converge to a solution, we need a
parameter that can robustly represent the value of $W_\text{in}$. While directly
controlling the value of $W_\text{in}$ or $r_\text{in}$ seems to be the most
straightforward and intuitive, we empirically find that fixing those quantities
during iterations often led to slow convergence or development of numerical
instabilities. We attempt to give a plausible explanation for this in
Sec.~\ref{sec:results}.

Alternatively, we parameterize the `depth' of the disk in terms of a parameter
$w \in [0, 1]$ defined as
\begin{equation} \label{eq:disk depth parametrization}
    W_\mathrm{in} = (1-w) W_\mathrm{center} + w W_c
\end{equation}
where the upper bound $W_c$ is defined by
\begin{equation} \label{eq:disk depth upper bound}
W_c \equiv
    \begin{cases}
    \max[ W_\mathrm{cusp}, \min(W(r=r_e), 0) ], & \text{if cusp exists} \\[1ex]
    \min[ W(r=r_e), 0 ], & \text{otherwise} \\
    \end{cases}
\end{equation}
The disk depth parameter\footnote{The $\delta$ parameter defined and used in
\cite{Corvino2009,Giacomazzo:2012bw} has the same role, despite a slightly
different form of definition.} $w$ is fixed during iterations and can be used to
control the mass of the disk. Having $w> 1$ may result in either an open
(infinite) disk or the disk being overfilled and exchanging pressure gradient
with the neutron star. We do not consider such complications here and only
consider equilibrium solutions with two separate fluid bodies. We still allow
some special cases where the surfaces (at which $\log H=0$) of the neutrons star
and the disk exactly touch each other, since such a configuration does not
disturb each of their hydrodynamic equilibria: see Sec.~\ref{sec:results} for
examples.

\begin{table*}
\centering
\caption{Convergence test on the neutron star-disk equilibrium model with the
    maximum rest energy density $\rhomax = 1.0\e{-3}$, neutron star axis ratio
    $r_p/r_e=0.60$, constant angular momentum disk ($a=0.0$) with $\kappa=7.00$,
    and the disk depth $q_w=0.90$. Each column shows the coordinate radius of
    the neutron star $r_e$, the angular velocity of the neutron star $\Omega_0$,
    rest mass of the neutron star $M\ns$, rest mass of the disk $M\disk$, and
    the total angular momentum of disk $J\disk$ for different grid resolutions.}
\label{tab:code convergence}
\begin{tabular}{ ccccccc }
\hline
    Grid resolution $(N_s \times N_\mu)$ & $r_e$ & $\Omega_c$ ($\e{-2}$) &
    $M\ns$ & $M\disk$ ($\e{-1}$) & $J\disk$ ($\e{-1}$)\\
\hline
    $401 \times 201$ & 12.098797070 & 2.3181343763 & 1.6639595312 & 1.0158858833
    & 7.0040203199 \\
    $801 \times 401$ & 12.099753746 & 2.3180761808 & 1.6641761685 & 1.0106603325
    & 6.9679129641 \\
    $1601 \times 801$ & 12.099821357 & 2.3180895689 & 1.6642203837 & 1.0097617642
    & 6.9617026170 \\
\hline
    \text{Convergence order} & 3.8 & 2.1 & 2.3 & 2.5 & 2.5 \\
\hline
\end{tabular}
\end{table*}

\subsection{Grid structure and discretization}

While the original KEH scheme used a truncated computational domain with a
finite radial extent, it is possible to compactify the spatial domain $r\in[0,
\infty)$ into a finite interval by introducing an appropriate coordinate
transformation \citep{Cook1992}. We define the radial coordinate variable $s(r)
\in [0, s_\mathrm{max}]$ as
\begin{equation}\label{eq:coordinate map}
    \frac{r}{r_e}=
    \begin{cases}
        s & \text{if} \quad 0 \leq s \leq s_0         \\
        s_0 + \sigma \tan \left( \frac{\pi}{2}\frac{s-s_0}{s_\mathrm{max}-s_0}
        \right)
          & \text{if} \quad s_0 < s \leq s_\text{max}
    \end{cases}
\end{equation}
with $s_0=1.0$ and $s_\mathrm{max} = 4.0$. The factor $\sigma$ is chosen as
\begin{equation}
    \sigma = \frac{2}{\pi}(s_\text{max}-s_0)
\end{equation}
in order to make the coordinate derivative $dr/ds$ continuous at the matching
point $s=s_0$. Compared to the coordinate transformation $r/ r_e = s / (1-s)$
introduced by \cite{Cook1992}, our radial coordinate map in
Eq.\eqref{eq:coordinate map} places a larger portion of grid points to the outer
region $r>r_e$ to resolve the structure of the disk.

Grid points are uniformly spaced in terms of the radial coordinate variable $s$
and the angular coordinate variable $\mu=\cos\theta$. For integrating field
equations we use the summation cutoff $L_\text{max} = 20$ in the expansion
series of Green's functions; see Eq.~(33)-(35) of \cite{Komatsu1989a}.  We use
the second-order discretization of spatial derivatives.

\subsection{Code tests}
\label{sec:code tests}

To validate our code for the rotating neutron star, we compare the total rest
(baryon) mass ($M_0$)
\begin{equation}
\begin{split}
M_0 & = \int \rho_0 u^t \sqrt{-\det g \,} d^3x, \\
    & = \int \frac{\rho_0 }{\sqrt{1-V^2}} e^{2\alpha + (\gamma - \varrho)/2}
            r^2 \sin\theta dr d\theta d\phi
\end{split}
\end{equation}
and angular momentum ($J$)
\begin{equation}
\begin{split}
    J & = \int T^t_\phi \sqrt{- \det g\,} d^3x \\
        & = \int \frac{(\rho_0 + \rho_i + p)V}{1-V^2}
        e^{2\alpha + (\gamma - \varrho)/2} r^3 \sin^2\theta dr d\theta d\phi
\end{split}
\end{equation}
with those from several available literature. Specifically, we compare our
results with those presented in \cite{Nozawa1998}, which conducted a detailed
comparison between three different numerical schemes
\citep{Komatsu1989a,Stergioulas1995,Bonazzola1993} devised for rotating neutron
star models. We check the values of $\bar{M}_0 \equiv K^{-1/2(\Gamma-1)} M_0$
and $\bar{J}\equiv K^{-1/(\Gamma-1)}J$ computed with the grid resolution $(N_s,
N_\mu) = (401,201)$, and our code shows a good agreement ($\lesssim$1\%) with
the results presented in \cite{Nozawa1998}. See Appendix~\ref{sec:detailed code
comparison} for detailed comparisons.

For models with a self-gravitating disk, we check the convergence of physical
quantities with increasing spatial grid resolutions. We consider a model with
neutron star parameters $(\rhomax, r_p/r_e) = (1.0\e{-3}, 0.60)$, disk rotation
parameters $(\kappa, a) = (7.00, 0.0)$, and the disk depth $q_w = 0.90$ as a
testing case.

Convergence test results are shown in Table~\ref{tab:code convergence}. We
increase the grid resolution by a factor of two, and measure the relative
convergence order with the formula $\mathcal{O} \equiv \log_2
(|q_\text{L}-q_\text{M}| / |q_\text{M}-q_\text{H}|) $, where $q_{\{\}}$ is a
physical quantity computed with the grid resolution L (low), M (medium), or H
(high). The measured order of convergence is consistent with the second-order
spatial discretization we used.

From Table~\ref{tab:code convergence}, it can be seen that the difference in
physical quantities between grid resolutions is larger for the disk (e.g.
$M\disk$) compared to those of the neutron star (e.g. $M\ns$). This lower
accuracy can be attributed to two reasons. First, in the KEH scheme, solving the
field equations is done by using Green's functions expansion at $r=0$, which has
a slower convergence of the series if $r \gtrsim r_e$. Second, the coordinate
map \eqref{eq:coordinate map} places radial grids with larger spacings in the
region $r/r_e > 1$, which leads to a larger truncation error.

\section{Results}
\label{sec:results}

\begin{figure*}
\centering
\includegraphics[width=\linewidth]{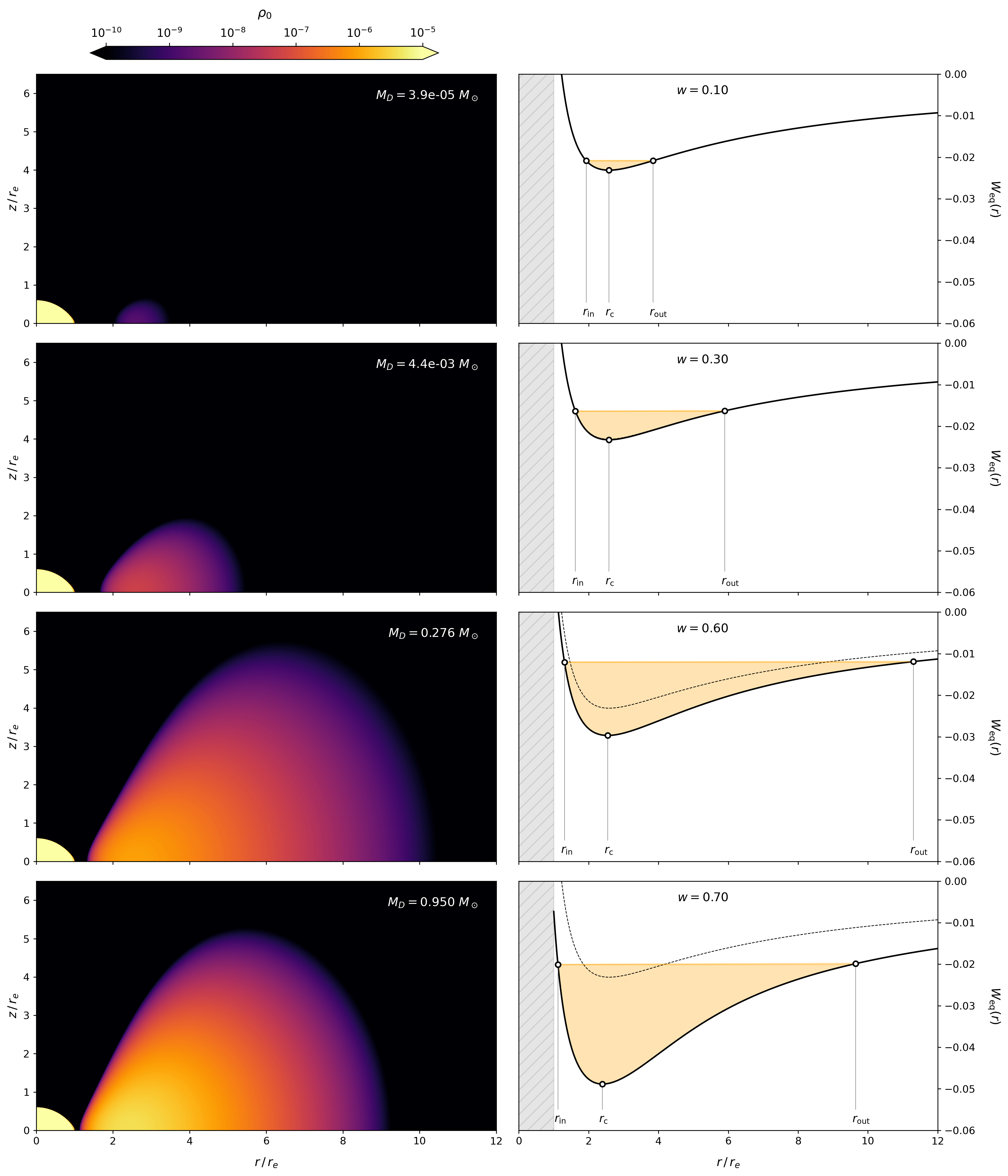}
\caption{Illustration of the computational procedure described in
    Sec.~\ref{sec:method outline} with the maximum rest mass density $\rhomax =
    1.0\e{-3}$, neutron star axis ratio $r_p/r_e=0.60$, disk rotation parameters
    $\kappa=7.78$ and $a=0.0$.
    We show rest mass density on the meridional plane (left panels) and the
    effective potential on the equatorial plane $W_\mathrm{eq}(r)$ (right
    panels) for increasing values of $w$ from top to bottom. By controlling the
    parameter $w$, rest mass of the disk $M\disk$ can be increased up to a
    critical configuration $M\disk=0.95\solarmass$ (bottom).
    In the right panels, we show the locations of the inner edge of the disk
    $r_\mathrm{in}$, center of the disk $r_\mathrm{c}$, and outer edge of the
    disk $r_\mathrm{out}$. The interior of the neutron star ($r<r_e$) is shown
    with a gray shade and the interior of the disk ($W_\mathrm{eq}(r) <
    W_\mathrm{in}$) is shown with an orange shade.}
\label{fig:single sequence}
\end{figure*}

\begin{table*}
\renewcommand{\arraystretch}{1.1}
\centering
\caption{Selected list of computed models and their parameters. $\rhomax$ is the
    maximum rest mass density, $r_p/r_e$ is the neutron star axis ratio, and
    $(\kappa, a)$ are rotation parameters in the angular momentum profile
    Eq.~\eqref{eq:disk rotation law} of the disk.
    Computed physical quantities are the rest mass of the neutron star $M\ns$,
    the coordinate radius of the neutron star $r_e$, the coordinate rotation
    period of the neutron star $T\ns = 2\pi / \Omega_0$, and the total rest mass
    of the disk $M\disk$. We also show $r_\text{in}$, $r_\text{center}$, and
    $r_\text{out}$, which are locations of the inner edge, center, and the outer
    edge of the disk on the equatorial plane in the unit of the neutron star
    coordinate radius $r_e$.
    For each set of the model parameters $(\rhomax, r_p/r_e, \kappa, a)$, we
    show in this table an equilibrium solution with the maximum disk mass with
    the outer edge of the disk not exceeding $r_\text{out} = 20r_e$.}
\label{tab:model lists}
\begin{tabular}{cc|cc|ccccccc|cc}
    \hline\hline
    $\rhomax$
    & $r_p/r_e$
    & $\kappa$ & $a$
    & $M\ns$
    & $r_e$
    & $T\ns$
    & $M\disk$
    & $r_\text{in}$
    & $r_\text{center}$
    & $r_\text{out}$
    & Name
    & Figure
    \\
    $[\si{\gram\per\cubic\cm}]$ & & & & [$\solarmass$] & [km] & [ms] & [$\solarmass$] & [$r_e$] & [$r_e$] & [$r_e$]
    \\
    \hline\hline
$6.17\e{14}$ & 0.999 & - & - & 1.354 & 12.8994 & 23.4173 & - & - & - & - & \\
    \cline{3-13}
& & 5.31 & 0.0 & 1.354 & 12.8985 & 25.0266 & 1.00$\e{-3}$ & 1.00 & 1.64 & 3.59
    & A1 & \ref{fig:r1_a999_a00:q008} \\
& & 5.52 & & 1.353 & 12.8882 & 172.513 & 1.67$\e{-2}$ & 1.00 & 1.86 & 6.18
    & A2 & \ref{fig:r1_a999_a00:q011} \\
& & 6.20 &  & 1.353 & 12.8751 & 118.016 & 5.97$\e{-2}$ & 1.39 & 2.62 & 10.9
    & A3 & \ref{fig:r1_a999_a00:q021} \\
& & 7.30 &  & 1.353 & 12.8525 & 44.2607 & 0.199 & 2.04 & 3.97 & 19.6
    & A4 \\ 
    \cline{2-13}
    & 0.600 & - & - & 1.665 & 17.9244 & 1.3312 & - & - & - & - & \\
    \cline{3-13}
& & 6.62 & 0.0 & 1.665 & 17.9197 & 1.3316 & 5.66$\e{-3}$ & 1.00 & 1.67 & 4.06
    & B1 & \ref{fig:r1_a60_a00:q009} \\
& & 7.40 &  & 1.661 & 17.5401 & 1.3566 & 0.760 & 1.00 & 2.14 & 9.14
    & B2 & \ref{fig:r1_a60_a00:q017} \\
& & 7.88 &  & 1.662 & 17.4585 & 1.3617 & 1.02 & 1.14 & 2.46 & 9.65
    & B3 & \ref{fig:r1_a60_a00:q022} \\
& & 9.62 &  & 1.665 & 17.8316 & 1.3375 & 0.398 & 2.17 & 4.23 & 18.7
    & B4 \\
& & 5.39 & 0.4 & 1.665 & 17.9244 & 1.3312 & 1.68$\e{-5}$ & 1.72 & 4.50 & 11.5
    & C1 & \ref{fig:r1_a60_a04:q002} \\
& & 5.42 &  & 1.665 & 17.9244 & 1.3312 & 2.44$\e{-4}$ & 1.93 & 5.11 & 19.2
    & C2 & \ref{fig:r1_a60_a04:q003} \\
& & 5.51 &  & 1.665 & 17.9244 & 1.3312 & 7.09$\e{-5}$ & 3.54 & 6.98 & 19.6
    & C3 & \ref{fig:r1_a60_a04:q006} \\
\hline
$1.23\e{15}$ & 0.600 & - & - & 2.024 & 14.2507 & 0.943 & - & - & - & - & \\
    \cline{3-13}
& & 6.95 & 0.0 & 2.024 & 14.2450 & 0.944 & 7.40$\e{-3}$ & 1.00 & 1.71 & 4.14
    & D1 & \ref{fig:r2_a60_a00:q010} \\
& & 7.48 &  & 2.022 & 14.0351 & 0.954 & 0.419 & 1.00 & 2.08 & 8.60
    & D2 & \ref{fig:r2_a60_a00:q017} \\
& & 7.70 &  & 2.022 & 13.9675 & 0.958 & 0.586 & 1.07 & 2.23 & 9.04
    & D3 & \\
& & 8.22 &  & 2.023 & 14.1591 & 0.948 & 0.242 & 1.46 & 2.79 & 11.0
    & D4 & \ref{fig:r2_a60_a00:q027} \\
& & 5.42 & 0.4 & 2.024 & 14.2507 & 0.943 & 1.05$\e{-7}$ & 2.81 & 5.45 & 8.86
    & E1 & \ref{fig:r2_a60_a04:q001} \\
& & 5.44 &  & 2.024 & 14.2507 & 0.943 & 1.46$\e{-5}$ & 2.49 & 6.35 & 15.2
    & E2 & \ref{fig:r2_a60_a04:q002} \\
& & 5.50 &  & 2.024 & 14.2507 & 0.943 & 3.65$\e{-5}$ & 4.03 & 7.98 & 19.6
    & E3 & \ref{fig:r2_a60_a04:q004} \\
\hline\hline
\end{tabular}
\end{table*}

Our numerical model requires five parameters listed in Table~\ref{tab:model
parameters}. Rather than varying all those parameters in an equal manner, we
consider a relatively limited number of neutron star configurations to focus on
exploring various species of the self-gravitating disk.

We consider the maximum rest mass density\footnote{Our code implementation does
not enforce $\rhomax$ to be the rest mass density at the center of the neutron
star, but we did not find any cases in which the maximum value of the rest mass
density appears other than at the grid origin $r=0$. This is no longer true for
neutron star models with more realistic, differential rotation profiles
\cite[e.g.][]{Uryu2017}.}
$\rhomax=1.0\e{-3}$ (which equals $6.17\e{14}\si{\gram\per\cubic\cm}$ in the
physical unit) and consider two different neutron star axis ratios: slowly
rotating ($r_p/r_e = 0.999$) and rapidly rotating ($r_p/r_e = 0.60$) cases.
Additionally, we also consider a rapidly rotating case with a higher central
density $(\rhomax, r_p/r_e) = (2.0\e{-3}, 0.60)$ to qualitatively study the
effect of neutron star mass on disk configurations. On rotation of the disk, we
only consider the prograde rotation ($\kappa > 0$) with constant ($a=0$) and
sub-Keplerian ($a=0.4$) angular momentum distribution on the equatorial plane.

A simple $K=100$, $\Gamma=2$ polytrope is used for modeling the neutron star.
For the matter constituting the equilibrium disk, we model it with the
relativistic electron degeneracy pressure; $K = 1.18\,Y_e^{4/3}$ with $Y_e =
0.5$ and $\Gamma=4/3$ \citep{Font2002,Daigne2004}.

All computations were performed with the grid resolution $(N_s, N_\mu) =
(801,401)$. The relative disk depth parameter $w$ is increased with a fixed
numerical increment $1/100$ during the construction of model sequences.

\subsection{A representative example}

Figure~\ref{fig:single sequence} illustrates our strategy of constructing a
sequence of equilibrium models for the parameters $\rhomax = 1.0\e{-3}$,
$r_p/r_e = 0.6$, $\kappa = 7.78$, $a=0.0$ following the computational procedure
described in Sec.~\ref{sec:method outline}. We fix the disk depth parameter $w$
in each set of iterations to converge to a single equilibrium model, then
proceed to the next solution with a larger disk mass by using a slightly
increased $w$ that is fixed in the next set of iterations. We show four
equilibrium models with increasing values of $w$ in Figure~\ref{fig:single
sequence}, including a critical model (at the bottom of the figure) beyond which
our code failed to converge. We note that open disks with infinite mass shown in
\citep{Font2002} no longer exist for the self-gravitating cases.

Effects of the self-gravity of the disk can be clearly seen through the
nonlinear distortion of $W_\text{eq}(r)$ curves, shown in the right panels of
Figure~\ref{fig:single sequence}. The potential dip gets deepened within the
interior of the disk, which is a clear indication of the disk self-gravity being
non-negligible compared to the background fields from the neutron star only. The
center of the disk $r_c$ remains almost unchanged until it is eventually shifted
slightly inward when the disk mass is large enough ($M\disk \gtrsim 0.5$). The
outer edge of the disk $r_\text{out}$ rapidly increases at a lower disk mass,
while it turns around and starts to move inward once the disk mass is
sufficiently large. Beyond this turning point of $r_\text{out}$, the radial
extent and thickness of the disk do not change significantly, or even shrink in
size (bottom panels of Figure~\ref{fig:single sequence}).

With the example presented in Figure~\ref{fig:single sequence}, we now revisit
the issue of parametrizing disk mass discussed in Sec.~\ref{sec:parametrizing
disk mass}. From the top and bottom panels of Figure~\ref{fig:single sequence},
it can be seen that the disk in both models, despite their large difference in
mass, shows approximately the same values of $W_\text{in}\approx -0.02$. This
degeneracy indicates that a single value of $W_\text{in}$ does not always
correspond to a unique configuration of the disk, showing its inadequacy as a
parameter for controlling the mass of the disk. Location of the inner edge of
disk $r_\text{in}$ exhibits a monotonic decrease with increasing disk mass, but
a steep slope of $W_\text{eq}(r)$ developing near $r\rightarrow r_e$ makes it
numerically challenging to find an optimal size of increment that can keep
iteration processes stable. In these regards, the method of parametrization
\eqref{eq:disk depth parametrization} shows less problematic behavior and
provides a more robust way to control the disk mass in actual computations.

\subsection{Equilibrium Models}

Table~\ref{tab:model lists} summarizes the parameters of and the resulting
physical quantities from a selected list of models.

Some models exhibit a radially extended disk with its outer edge reaching up to
$\sim$200km. As discussed in Sec.~\ref{sec:code tests}, the accuracy of the
result can be significantly lost if the radial extent of the disk is too large.
The radial grid resolution $N_s=801$ used in computation, in combination with
the radial coordinate map \eqref{eq:coordinate map} used in this work, places
only $\sim$40 grid points within $r\in(20r_e,\infty)$, putting the validity of a
numerical solution in question. For these reasons, models with $r_\text{out} >
20 r_e$ are discarded from the results.

Each of the entries in Table~\ref{tab:model lists} is understood as that a
continuous sequence of models can be obtained up to the value presented in the
table, at which the model is either at a critical rotation or the outer edge of
the disk reached $r_\text{out} \lesssim 20 r_e$. For instance, the entry B2
(corresponding to Figure~\ref{fig:r1_a60_a00:q017}) indicates that all
configurations with $M\disk \leq 0.76M_\odot$ could have been obtained with the
given parameters $\rhomax=1.0\e{-3}$, $r_p/r_e=0.60$, $\kappa=7.40$, and
$a=0.0$.

Several generic trends in physical quantities can be observed from
Table~\ref{tab:model lists}, by comparing results with those without
self-gravity of the disk. As the mass of the disk gets larger, the rest mass of
the neutron star decreases by a small amount ($\lesssim 0.1\%$). The coordinate
radius of the neutron star is also decreased with a slightly larger amount. The
neutron star also shows slower rotation when the disk exists.

This can be qualitatively understood as the gravitational pull from the disk,
pointing radially outward on the equatorial plane, provides extra centrifugal
support to the rotating matter inside the neutron star. Therefore, with fixed
maximum rest mass density and axis ratio, the matter distribution of the neutron
star is changed into a configuration with a slower rotation, resulting in a
smaller mass, smaller size, and a slower rotation frequency. In particular, the
gravitational effect from the disk is quite noticeable in the cases with a
slowly rotating ($r_p/r_e = 0.999$) neutron star, significantly increasing its
rotation period while leaving the rest mass and the coordinate radius almost
unchanged. On the contrary, the effects on the rotation frequency of the neutron
star are marginal for rapidly rotating ($r_p/r_e = 0.60$) cases. Overall, these
trends found in our results are also consistent with the findings of
\cite{Nishida1992}.

We look into details of the numerical models with a constant angular momentum
disk (model A, B, D) in Sec.~\ref{sec:constant angular momentum disk}, then we
discuss the qualitative effect of higher disk rotation index $a$ (model C, E) in
Sec.~\ref{sec:non-constant angular momentum disk}.

\begin{figure*}
\centering
\setlength{\columnsep}{0.1em}
\begin{multicols}{3}
\subcaptionbox{$\kappa=5.31$, $M\disk = 1.00\e{-3}M_\odot$ \label{fig:r1_a999_a00:q008}}
{\includegraphics[width=\linewidth]{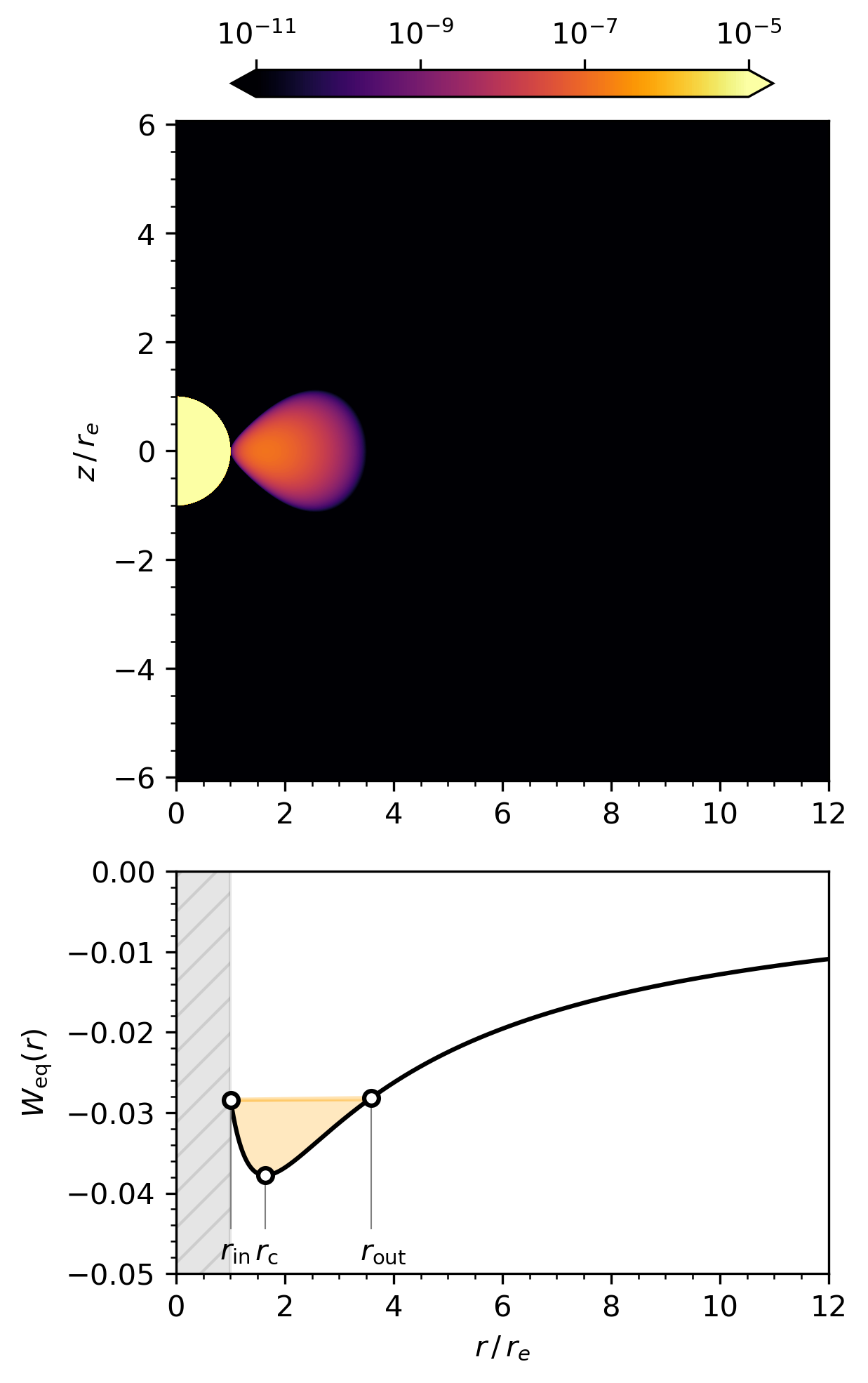}}
\par
\subcaptionbox{$\kappa=5.52$, $M\disk = 1.67\e{-2}M_\odot$ \label{fig:r1_a999_a00:q011}}
{\includegraphics[width=\linewidth]{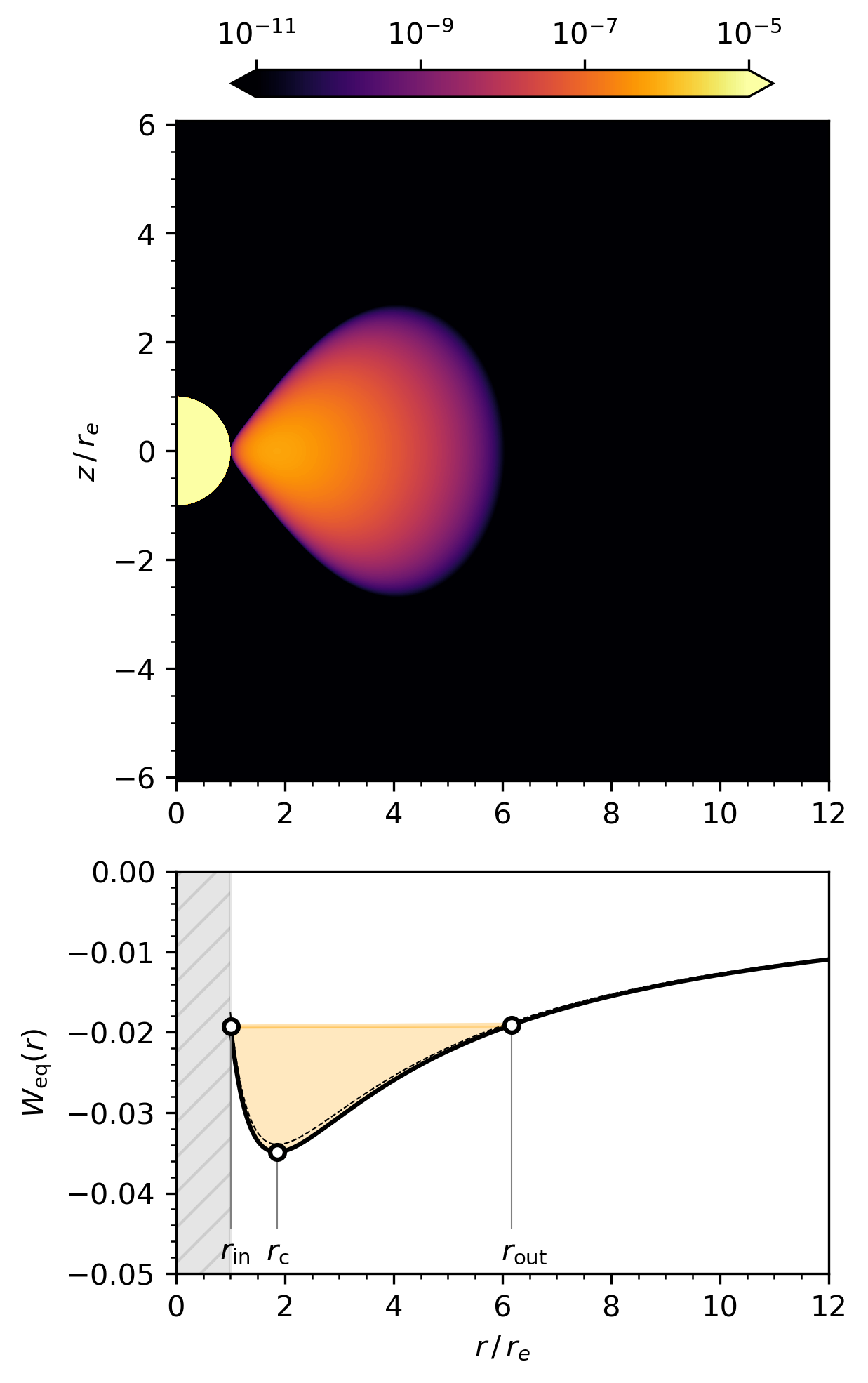}}
\par
\subcaptionbox{$\kappa=6.20$, $M\disk = 5.97\e{-2}M_\odot$ \label{fig:r1_a999_a00:q021}}
{\includegraphics[width=\linewidth]{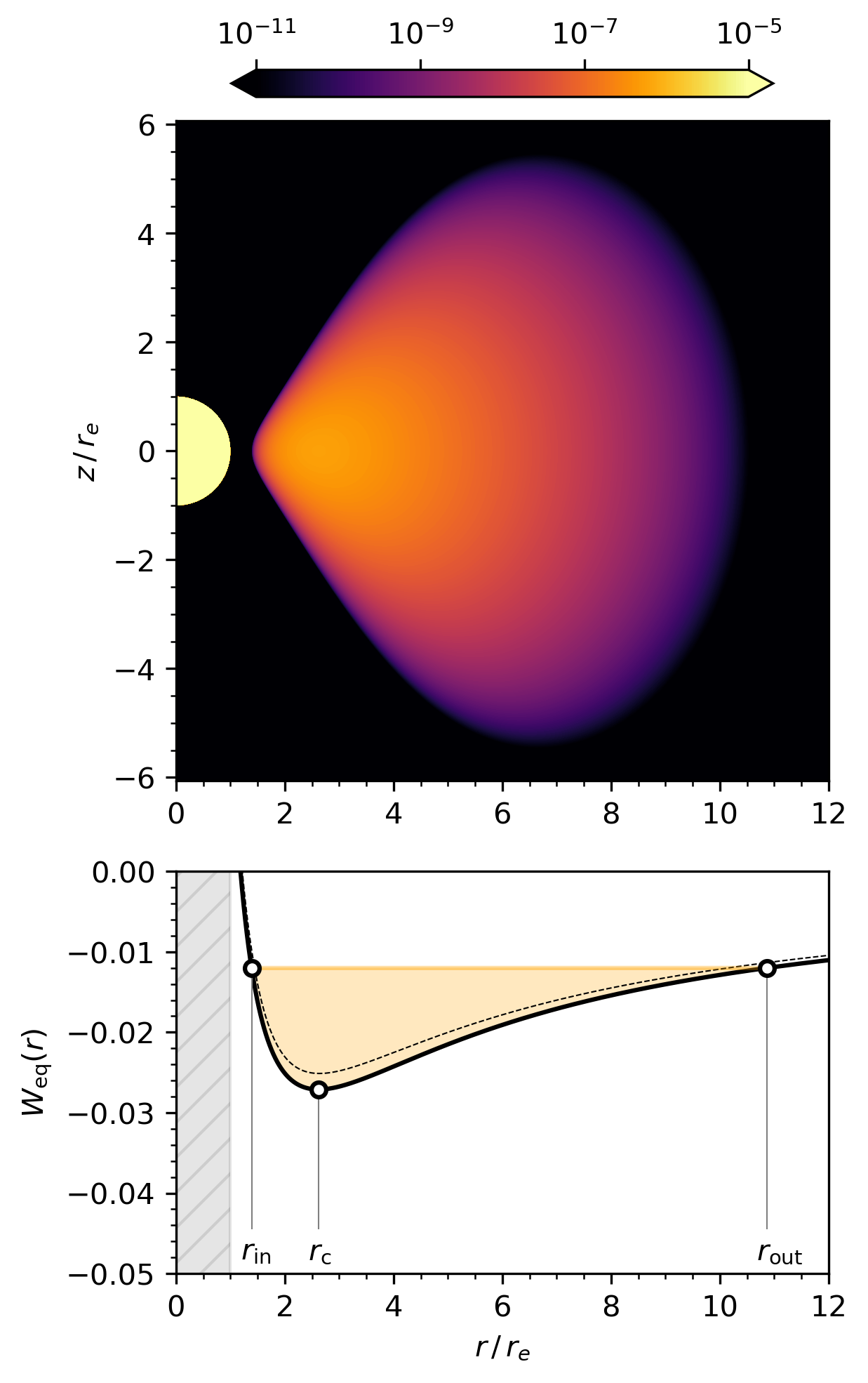}}
\end{multicols}
\vspace{-1em}
\caption{Self-gravitating star-disk models for $(\rhomax, r_p/r_e) = (1.0\e{-3},
    0.999)$ neutron star with a constant angular momentum disk ($a=0.0$).
    Equilibrium solutions with three values of $\kappa$ are shown: disk can be
    filled to $w= 1.0$ but self-gravity is still nearly negligible (left); the
    disk can be filled up but the effect of self-gravity becomes noticeable
    (center); the disk reaches a detached critical configuration ($r_e <
    r_\text{in}$) with non-negligible self-gravity (right).
    \emph{Upper panels}: Rest mass density $\rho_0$ on the meridional plane.
    \emph{Lower panels}: Effective potential $W_\text{eq}(r)$ is shown with a
    thick black line, and the potential without the disk self-gravity is shown
    with the dotted line. We mark the locations of the inner edge, center, and
    outer edge of the disk with circles, and display the interior of the disk
    with a yellow shade. Interior of the neutron star is shown with a grey
    shade.}
\label{fig:r1_a999_a00}
\end{figure*}

\begin{figure*}
\centering
\setlength{\columnsep}{0.1em}
\begin{multicols}{3}
\subcaptionbox{$\kappa=6.62$, $M\disk = 5.66\e{-3}M_\odot$ \label{fig:r1_a60_a00:q009}}
{\includegraphics[width=\linewidth]{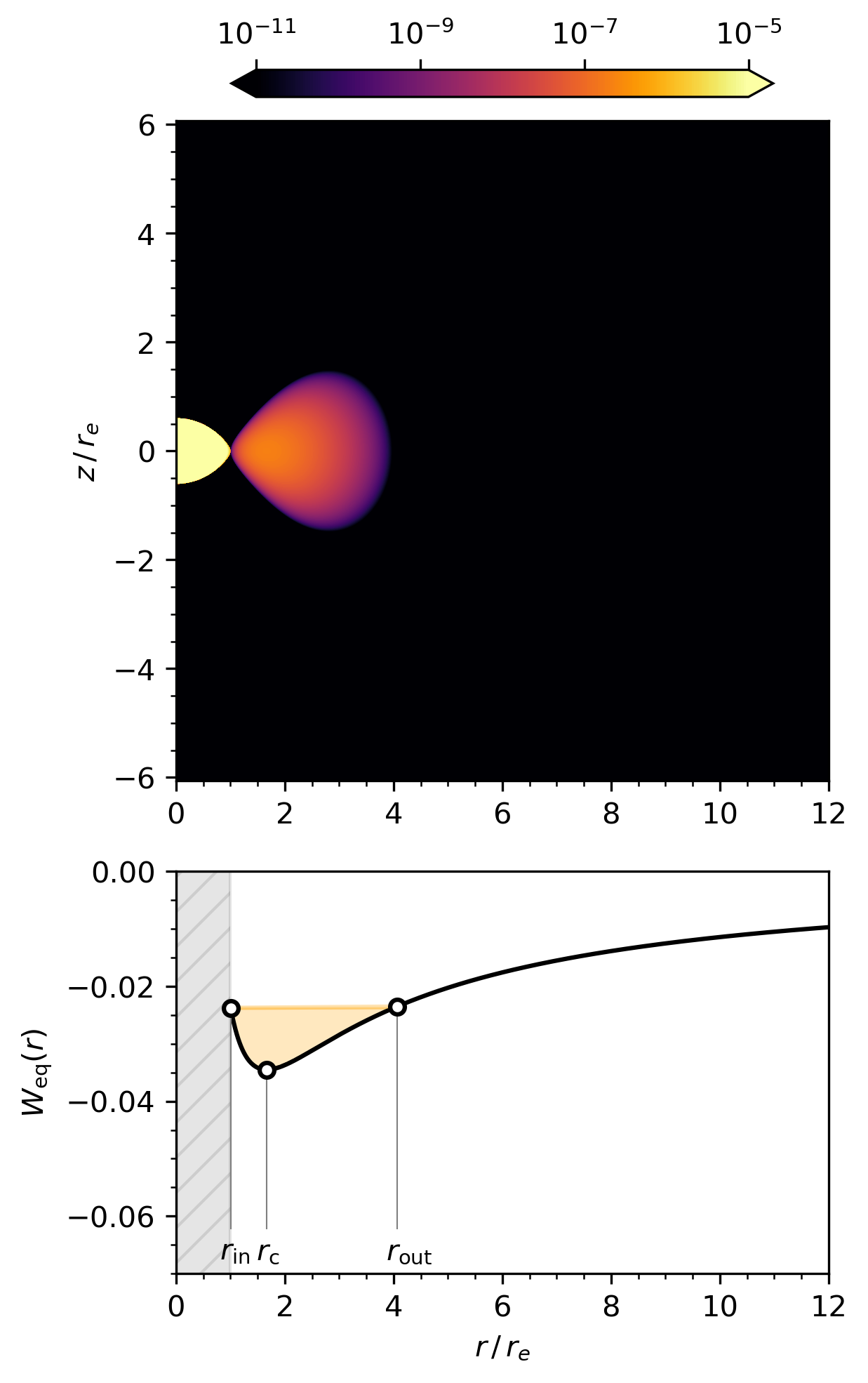}}
\par
\subcaptionbox{$\kappa=7.40$, $M\disk = 0.760M_\odot$ \label{fig:r1_a60_a00:q017}}
{\includegraphics[width=\linewidth]{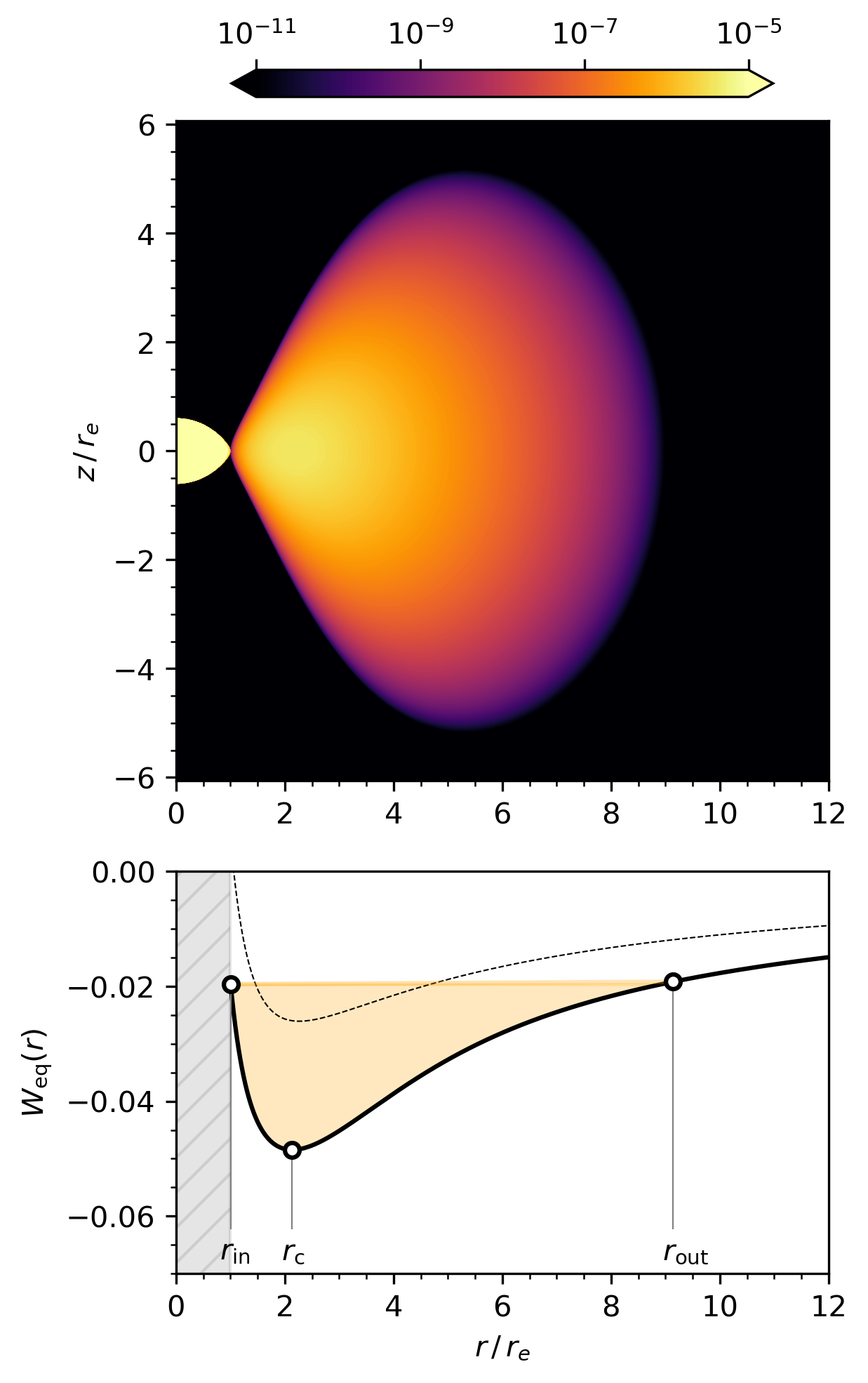}}
\par
\subcaptionbox{$\kappa=7.88$, $M\disk = 1.02M_\odot$ \label{fig:r1_a60_a00:q022}}
{\includegraphics[width=\linewidth]{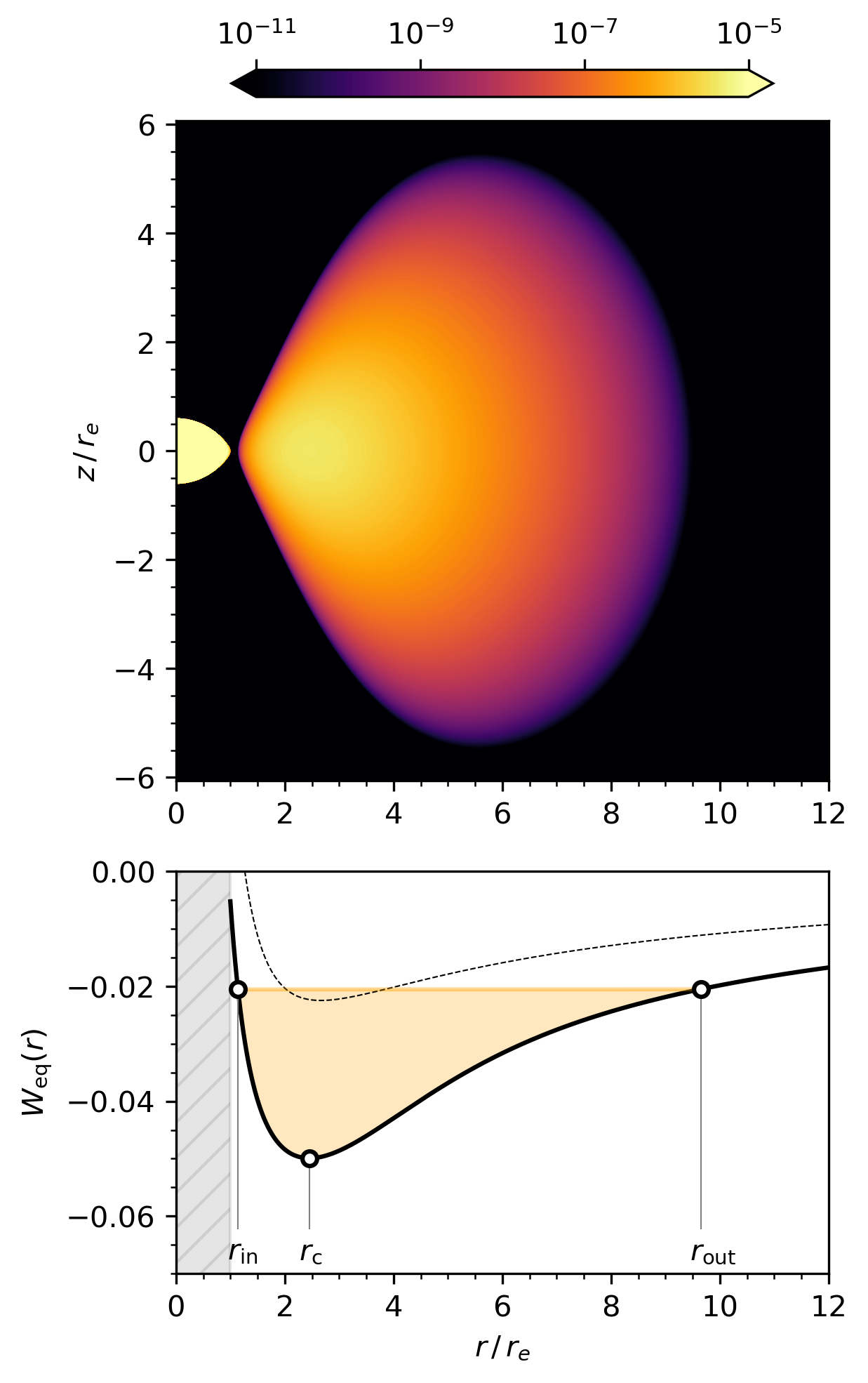}}
\end{multicols}
\vspace{-1em}
\caption{Same as Figure~\ref{fig:r1_a999_a00}, for rapidly rotating case $r_p/r_e
    = 0.60$.}
\label{fig:r1_a600_a00}
\end{figure*}

\begin{figure*}
\centering
\setlength{\columnsep}{0.1em}
\begin{multicols}{3}
\subcaptionbox{$\kappa=6.95$, $M\disk = 7.40\e{-3}M_\odot$ \label{fig:r2_a60_a00:q010}}
{\includegraphics[width=\linewidth]{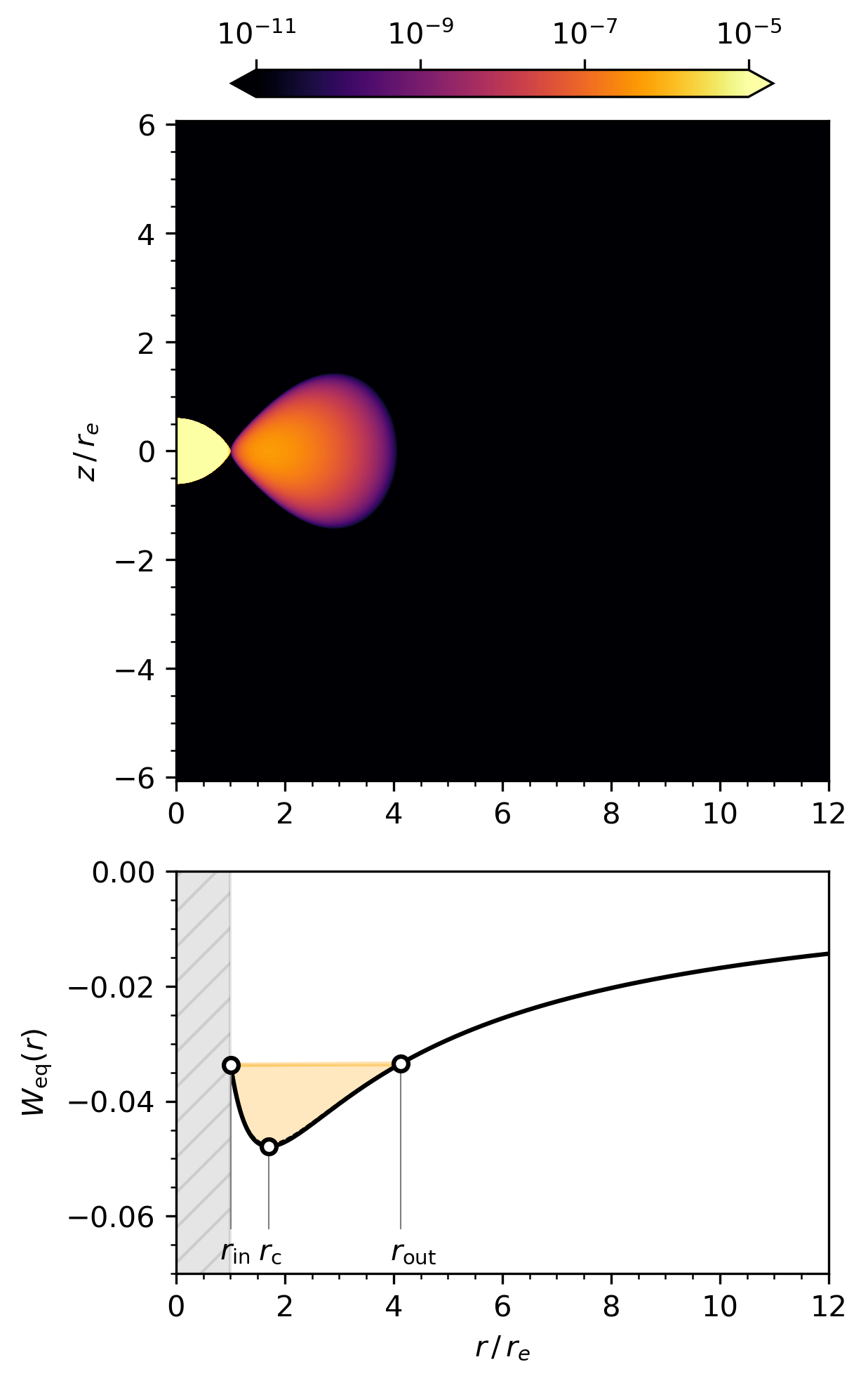}}
\par
\subcaptionbox{$\kappa=7.48$, $M\disk = 0.419M_\odot$ \label{fig:r2_a60_a00:q017}}
{\includegraphics[width=\linewidth]{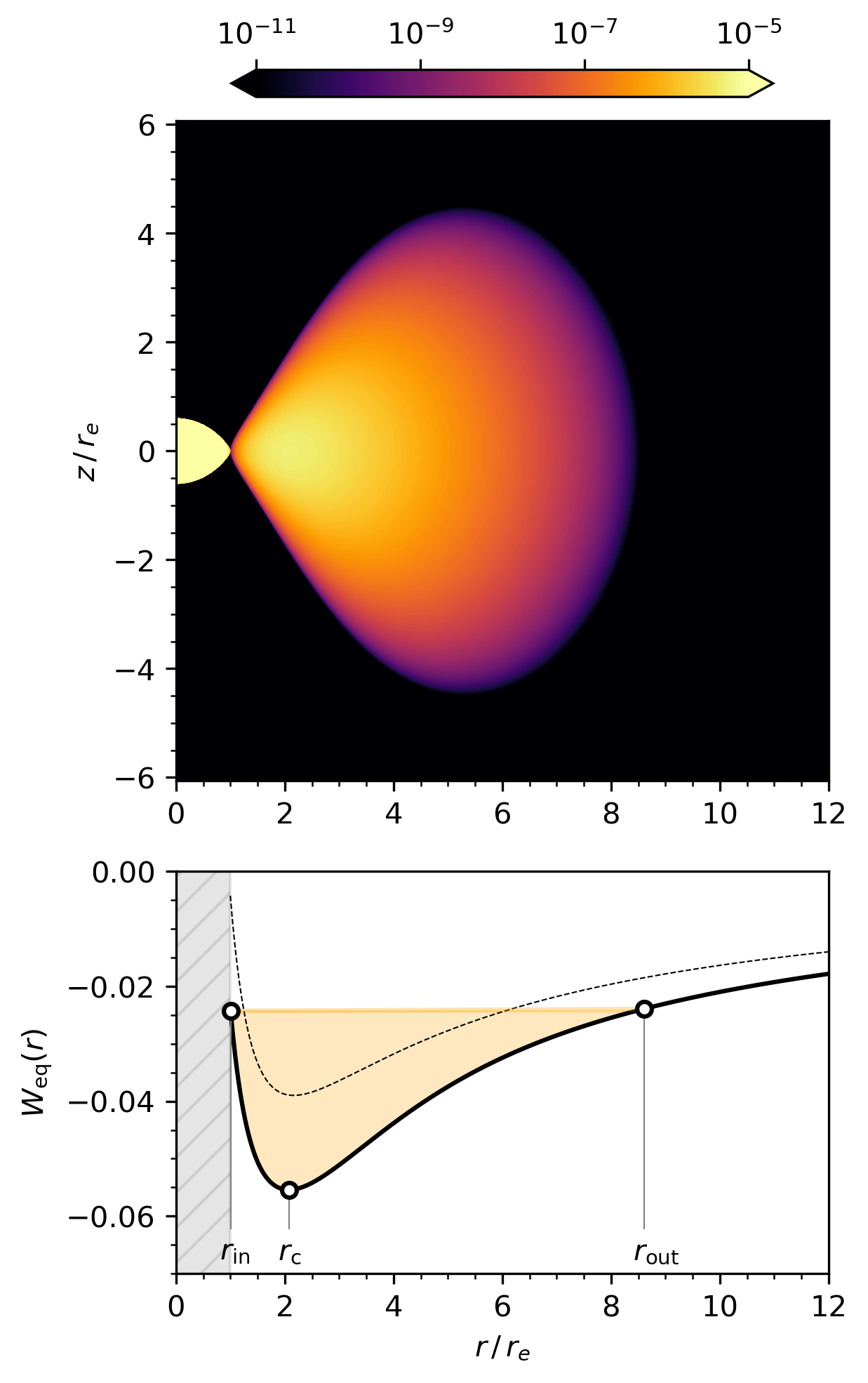}}
\par
\subcaptionbox{$\kappa=8.22$, $M\disk = 0.242M_\odot$ \label{fig:r2_a60_a00:q027}}
{\includegraphics[width=\linewidth]{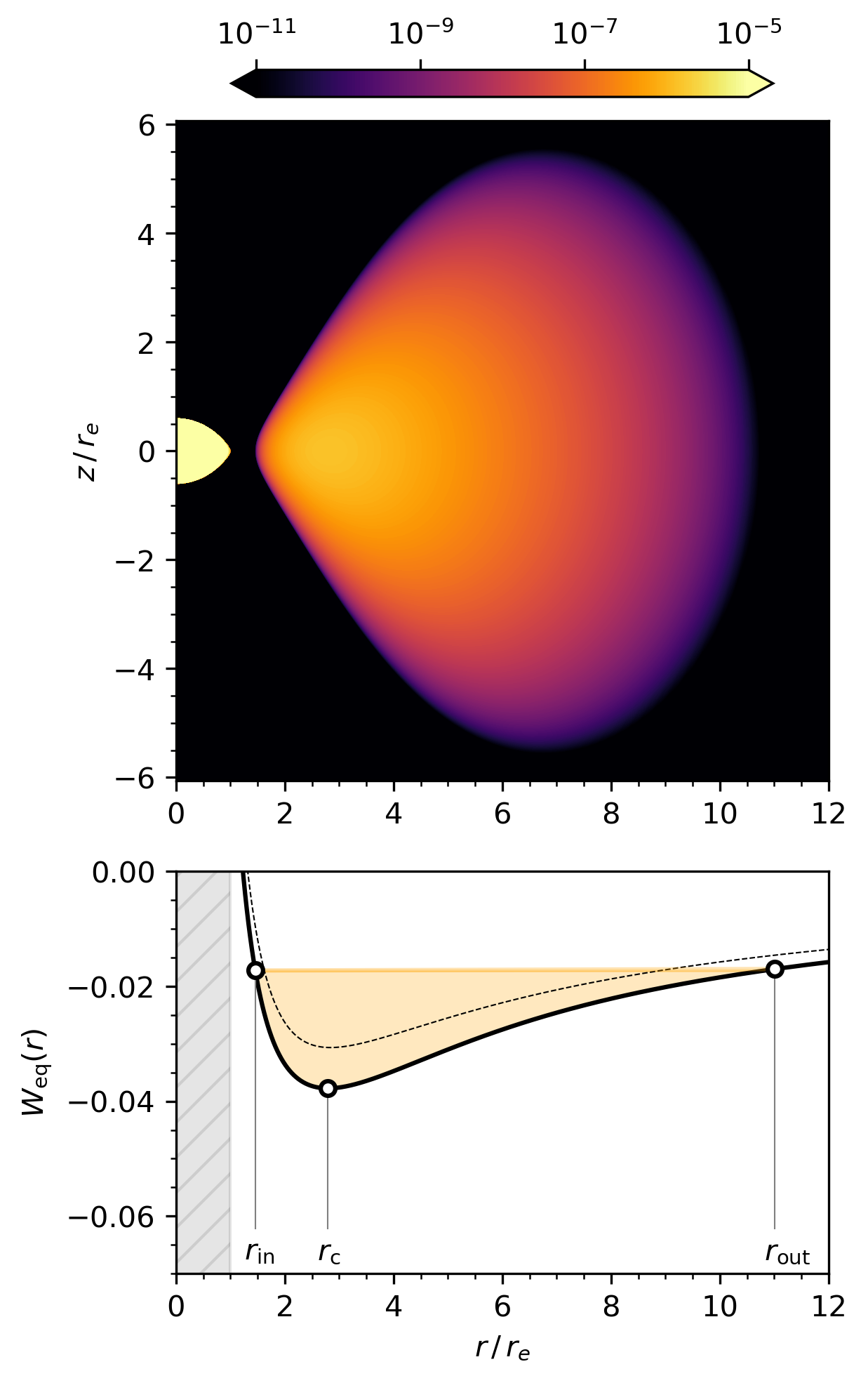}}
\end{multicols}
\vspace{-1em}
\caption{Same as Figure~\ref{fig:r1_a600_a00}, for a higher maximum rest mass
    density $\rhomax = 2\e{-3}$.}
\label{fig:r2_a600_a00}
\end{figure*}

\subsubsection{Constant angular momentum disk}
\label{sec:constant angular momentum disk}

Figure~\ref{fig:r1_a999_a00} shows the rest mass density on the meridional plane
for models involving a slowly rotating ($r_p/r_e=0.999$) neutron star with
$\rhomax=1.0\e{-3}$ and a constant angular momentum disk (models A1-A3). Three
selected values of $\kappa$ correspond to the configuration at which: (i) the
disk is located close and can be filled up to $w=1$, touching the neutron star
surface but its self-gravity is not significant (model A1,
Figure~\ref{fig:r1_a999_a00:q008}); (ii) same as the first case but the
self-gravity is non-negligible (model A2, Figure~\ref{fig:r1_a999_a00:q011});
(iii) the inner edge of the disk being detached from the neutron star surface,
cannot be fully filled ($w<1$), and the self-gravity is non-negligible (model
A3, Figure~\ref{fig:r1_a999_a00:q021}). In the last case, the model reaches a
critically rotating state at an upper limit $w_\text{max} = 0.56$.

The angular velocity of the neutron star is significantly affected by the
gravity from the disk; the neutron star in the model A2 shows more than seven
times slower rotation than the isolated case.
However, this should not be interpreted as that a rotating neutron star can
actually be `slowed down' to this amount by forming a disk around it, but needs
to be understood as the functional dependency between the neutron star axis
ratio $r_p/r_e$ and the rotation frequency $T\ns$ is very sensitive to the
influence from an external gravity in a slowly rotating regime.
Comparing models A1-A4, we see that the rotation speed of the neutron star is
most affected in model A2 even if model A4 has more than 10 times larger disk
mass. This indicates that the influence of the disk on the hydrodynamic
equilibrium of the neutron star is not solely dominated by the rest mass of the
disk, but its proximity to the neutron star surface also plays an important
role.

We show the same set of plots for a rapidly rotating neutron star
($r_p/r_e=0.60$) in Figure~\ref{fig:r1_a600_a00}. These models (B1-B4), with a
rapidly rotating neutron star, permits solutions with larger disk mass, even up
to $M\disk\approx 1.0$ (model B3, Figure~\ref{fig:r1_a60_a00:q022}); note the
large distortion of the effective potential $W_\text{eq}(r)$. Meanwhile, the
influence of the disk on the rotation period of the neutron star is small
($\lesssim 2\%$), implying that its relative contribution to the hydrodynamic
equilibrium of the neutron star is much smaller than the models with a slowly
rotating neutron star, as expected.

Models D1-D4 have a rapidly rotating neutron star with a higher maximum rest
energy density $\rhomax=2.0\e{-3}$; models D1, D2, and D4 are shown in
Figure~\ref{fig:r2_a600_a00}. The neutron star has a higher mass $M\ns \approx
2.0 M_\odot$. more compact radius $r_e \approx 14 \text{km}$, and shorter
rotation period $\leq 1\text{ms}$. This sequence yields models with the disk
mass being a bit smaller than the models B1-B4.

\begin{figure*}
\centering
\setlength{\columnsep}{0.1em}
\begin{multicols}{3}
\subcaptionbox{$\kappa=5.39$, $M\disk = 1.68\e{-5}M_\odot$ \label{fig:r1_a60_a04:q002}}
{\includegraphics[width=\linewidth]{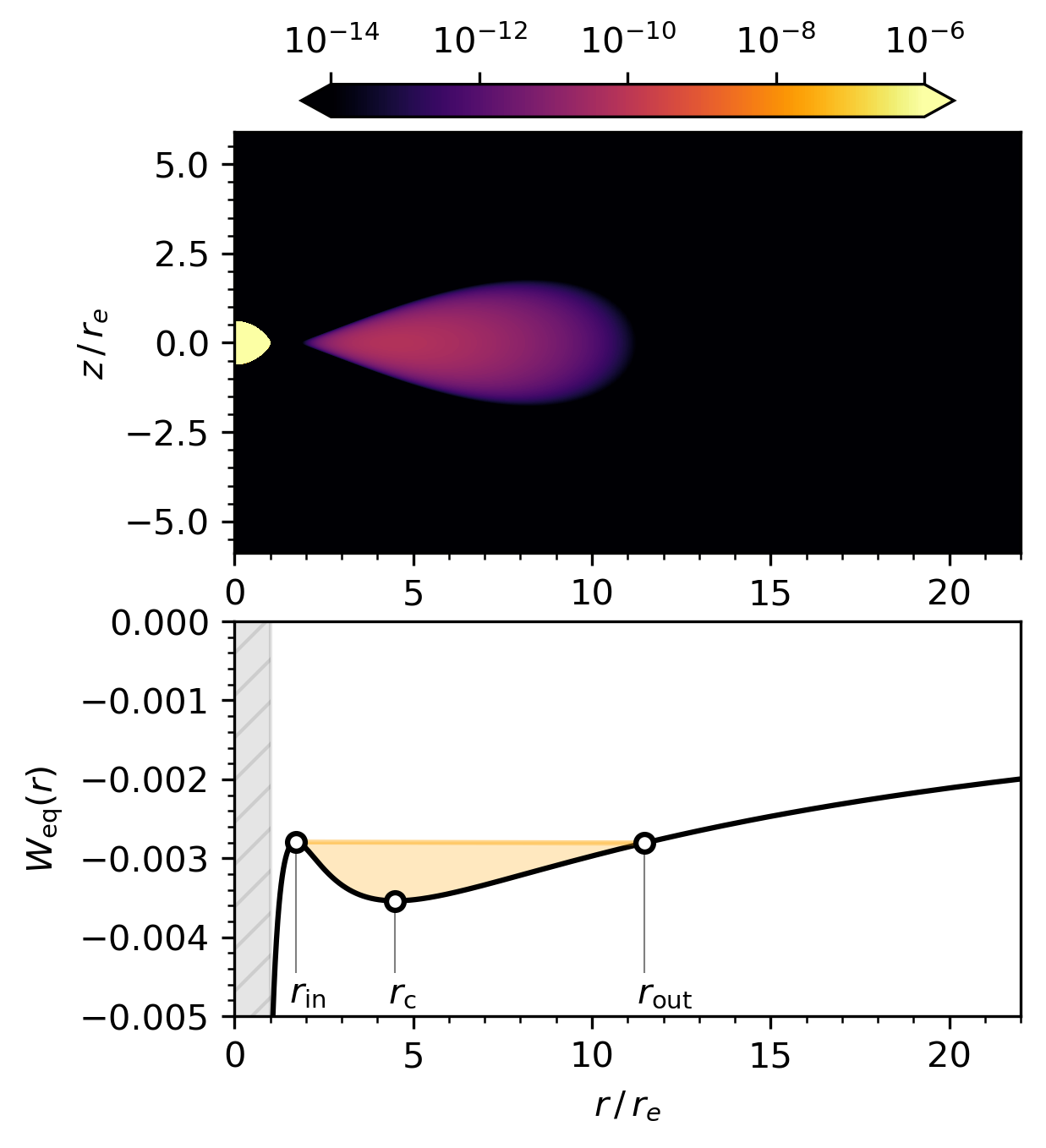}}
\par
\subcaptionbox{$\kappa=5.42$, $M\disk = 2.44\e{-4}M_\odot$ \label{fig:r1_a60_a04:q003}}
{\includegraphics[width=\linewidth]{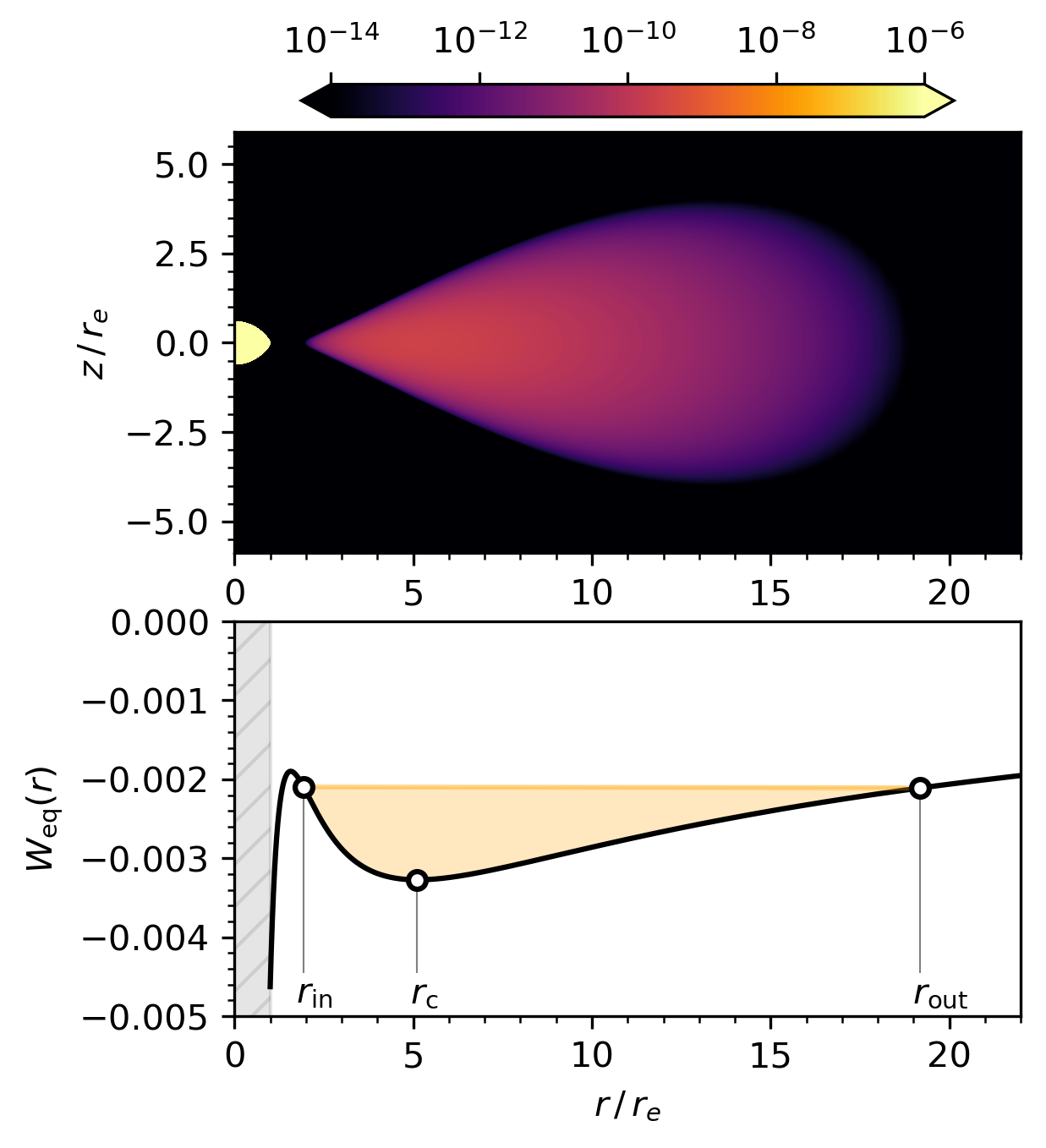}}
\par
\subcaptionbox{$\kappa=5.51$, $M\disk = 7.09\e{-5}M_\odot$ \label{fig:r1_a60_a04:q006}}
{\includegraphics[width=\linewidth]{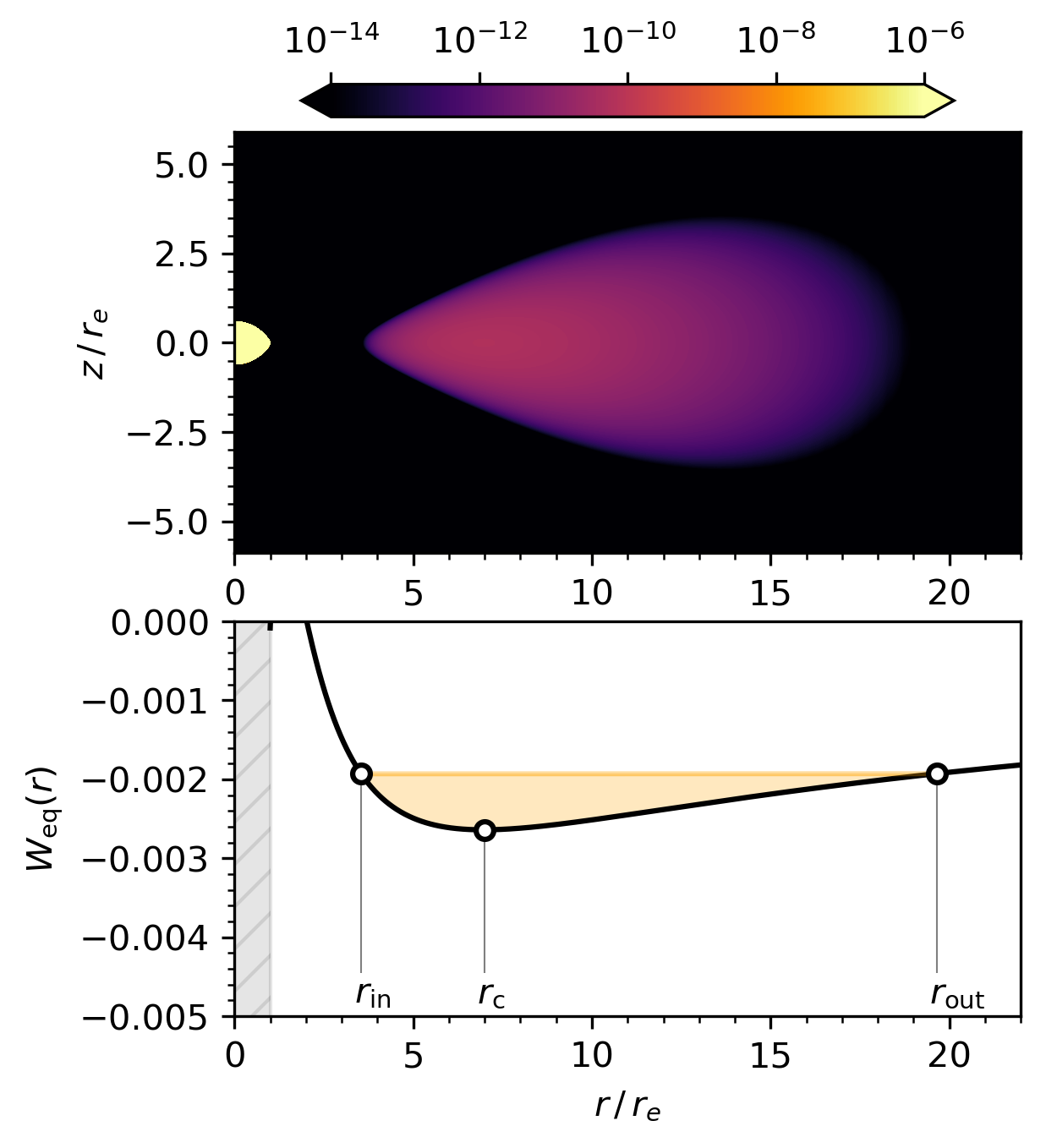}}
\end{multicols}
\vspace{-1em}
\caption{Non-constant angular momentum disk models. $\rhomax = 1.0\e{-3}$,
    $r_p/r_e = 0.60$, and $a=0.4$}
\label{fig:r1_a600_a04}
\end{figure*}

\begin{figure*}
\centering
\setlength{\columnsep}{0.1em}
\begin{multicols}{3}
\subcaptionbox{$\kappa=5.42$, $M\disk = 1.05\e{-7}M_\odot$ \label{fig:r2_a60_a04:q001}}
{\includegraphics[width=\linewidth]{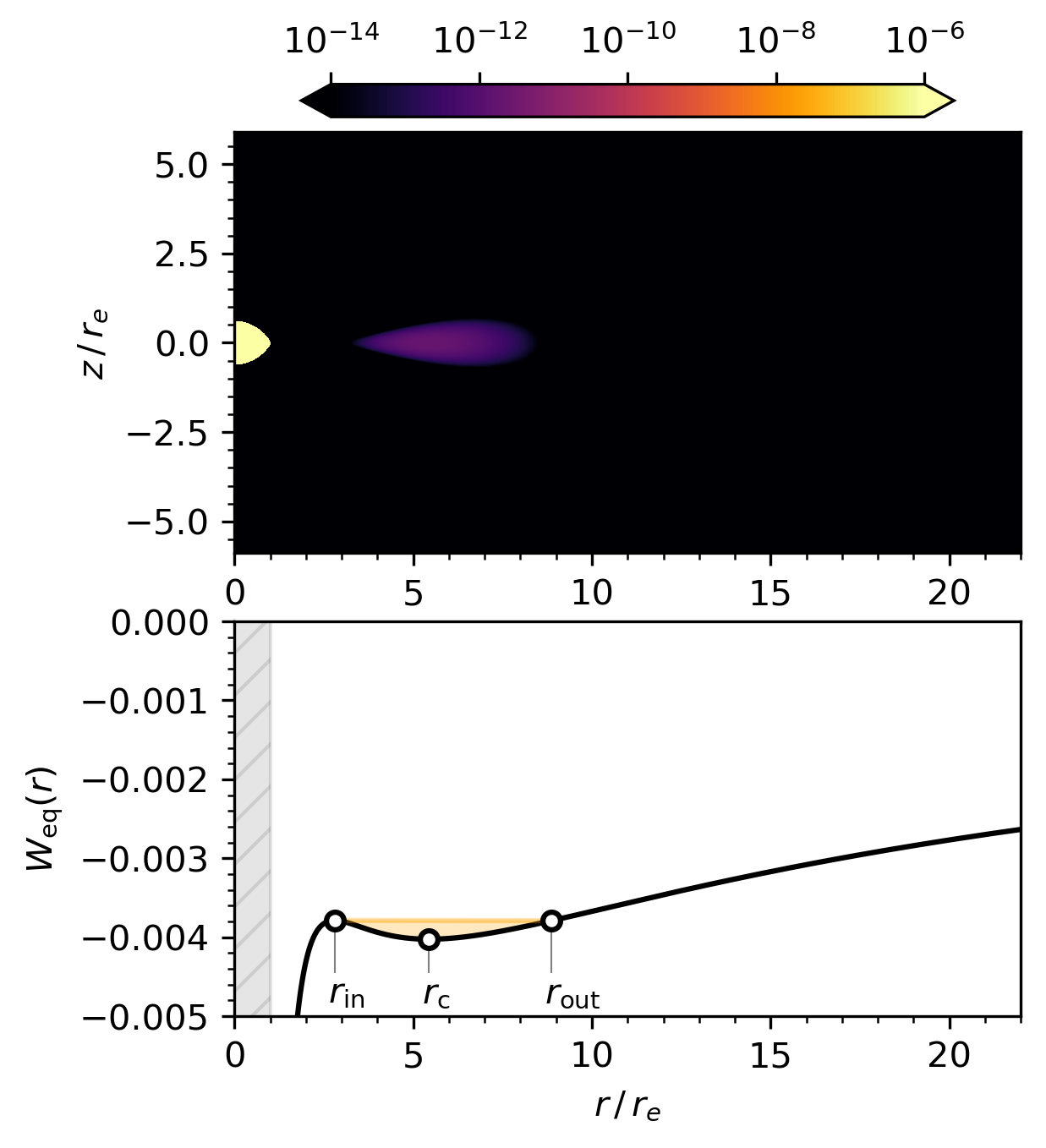}}
\par
\subcaptionbox{$\kappa=5.44$, $M\disk = 1.46\e{-5}M_\odot$ \label{fig:r2_a60_a04:q002}}
{\includegraphics[width=\linewidth]{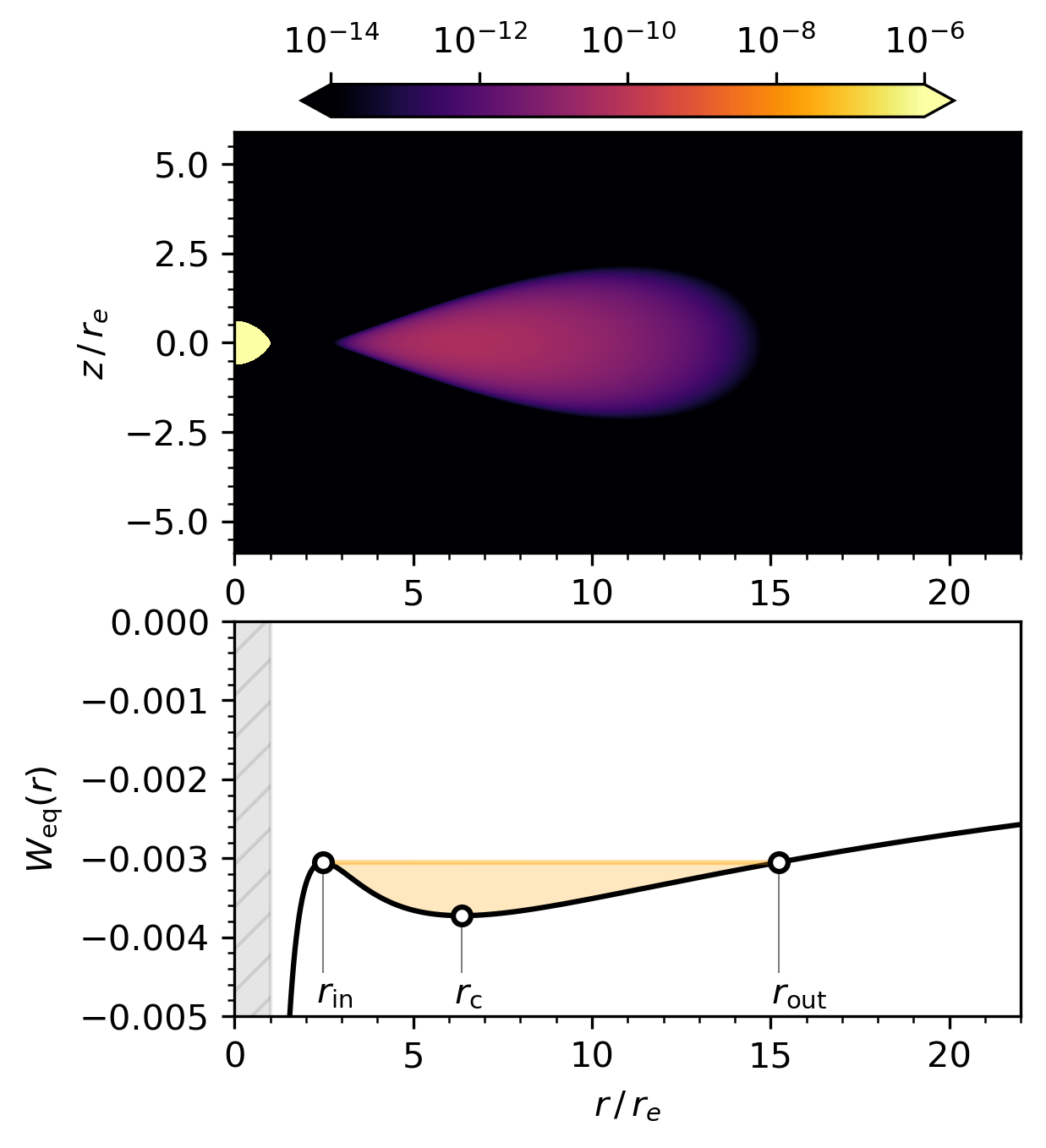}}
\par
\subcaptionbox{$\kappa=5.47$, $M\disk = 6.68\e{-5}M_\odot$ \label{fig:r2_a60_a04:q004}}
{\includegraphics[width=\linewidth]{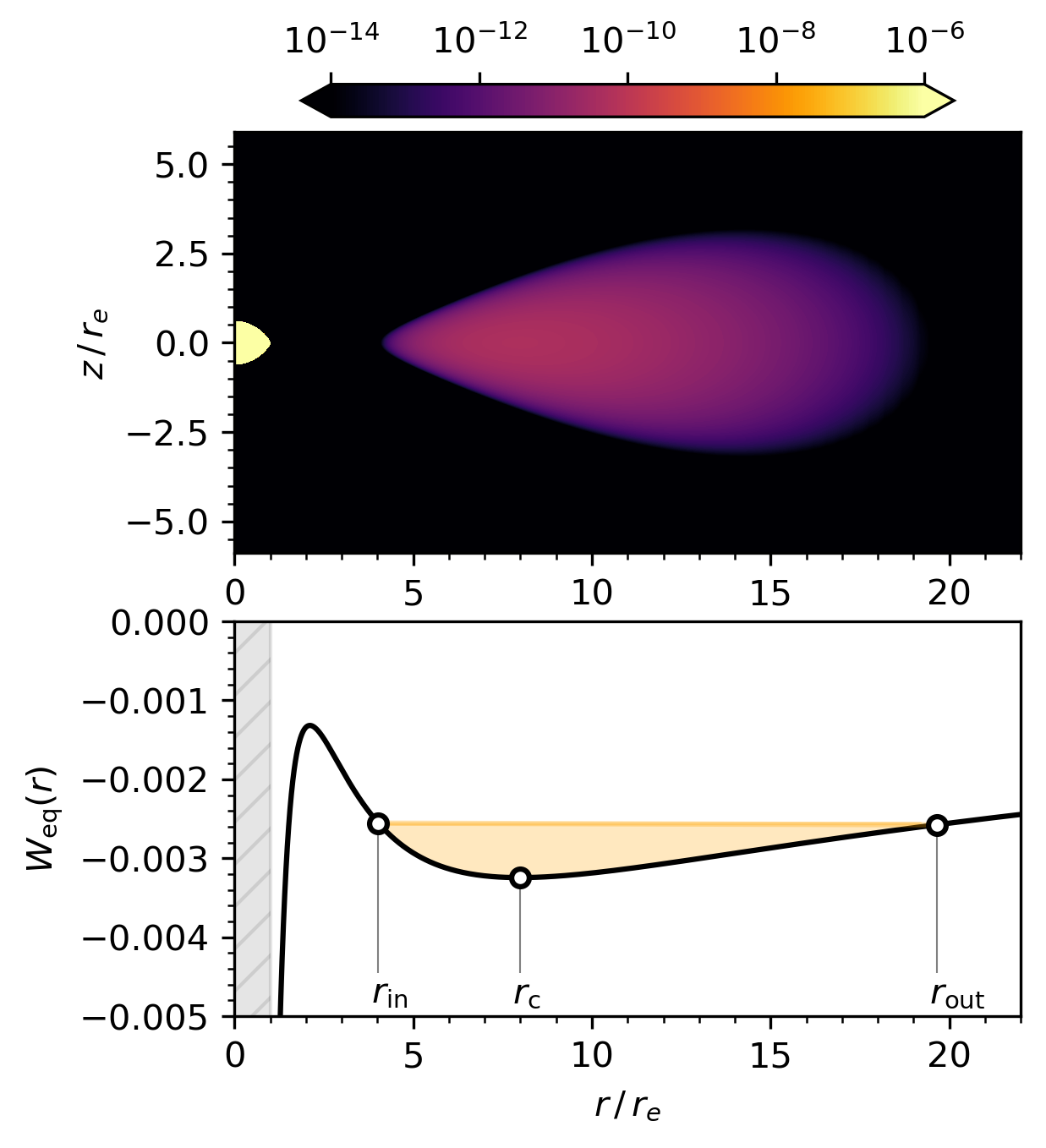}}
\end{multicols}
\vspace{-1em}
\caption{Same as Figure~\ref{fig:r1_a600_a04}, for a higher maximum rest mass
    density $\rhomax = 2.0\e{-3}$}
\label{fig:r2_a600_a04}
\end{figure*}

\subsubsection{Non-constant angular momentum disk}
\label{sec:non-constant angular momentum disk}

As discussed in Sec.~\ref{sec:bounds on disk parameters}, a higher power index
$a$ facilitates the development of a cusp in the effective potential
$W_\text{eq}(r)$, effectively prohibiting the disk with a large mass. Our
parameter study confirms that it is difficult to obtain a solution with a large
disk mass; all our results show $M\disk \lesssim 10^{-4} M_\odot$.

In Figure~\ref{fig:r1_a600_a04} (Figure~\ref{fig:r2_a600_a04}), we show the rest
mass density on the meridional plane and the effective potential of models C1-C3
(E1-E3). It can be seen that the cusp is present near the surface of the neutron
star, and any part of the disk with its effective potential exceeding the value
of $W_\text{cusp}$ is not able to sustain an equilibrium but will overflow and
accrete onto the neutron star. Note that the disk filling exactly up to the cusp
of the potential $W_\text{eq}(r)$ shows a sharp and thin inner edge (e.g.
Figure~\ref{fig:r1_a60_a04:q002}), which conforms with the boundary of the Roche
lobe.

In general, compared to the constant angular momentum disks, sub-Keplerian disk
models show more "shallow and flat" curves of $W_\text{eq}(r)$. With a high
value of $a$ close to the Keplerian ($a=0.5$) limit, the presence of the cusp
prevents the formation of a potential `dip' in the $W_\text{eq}$ curve. This has
two consequences in equilibrium solutions, namely that (1) the mass of the disk
is much smaller overall, and (2) the radial extent of the disk grows very fast
with increasing disk depth parameter $w$. We also note that the viable range of
$\kappa$ in which a disk can be constructed is much narrower than the constant
angular momentum cases (see Table~\ref{tab:model lists}).

\subsection{Comparisons with non-self-gravitating disks}
\label{sec:importance of self gravity}

\begin{figure}
\centering
\includegraphics[width=\linewidth]{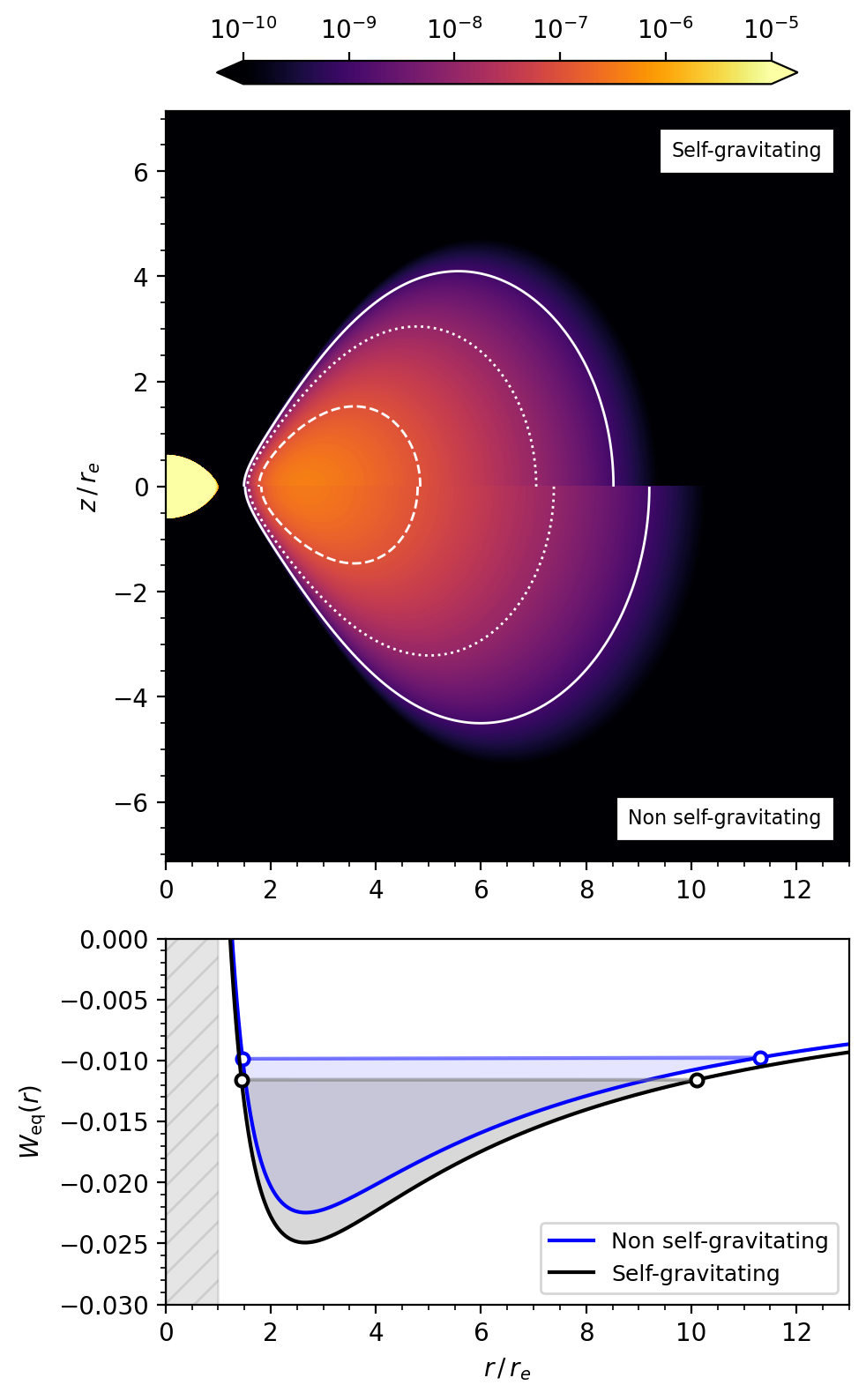}
\caption{Comparison between models with and without the disk self-gravity for
$\kappa = 7.88$, $M\disk = 0.101M_\odot$.
(Top) Rest mass density of the disk in self-gravitating and non-self-gravitating
cases. To better visualize the difference of the matter distribution between two
cases, we draw contour lines for $\rho_0=10^{-9}, 10^{-8}, 10^{-7}$ with solid,
dotted, and dashed white lines, respectively.
(Bottom) Effective potential $W_\text{eq}(r)$. The inner and outer edges of the
disk are shown with small circles, and the interior region of the disk is
displayed with colored shades.}
\label{fig:comparison low mass}
\end{figure}

\begin{figure}
\centering
\includegraphics[width=\linewidth]{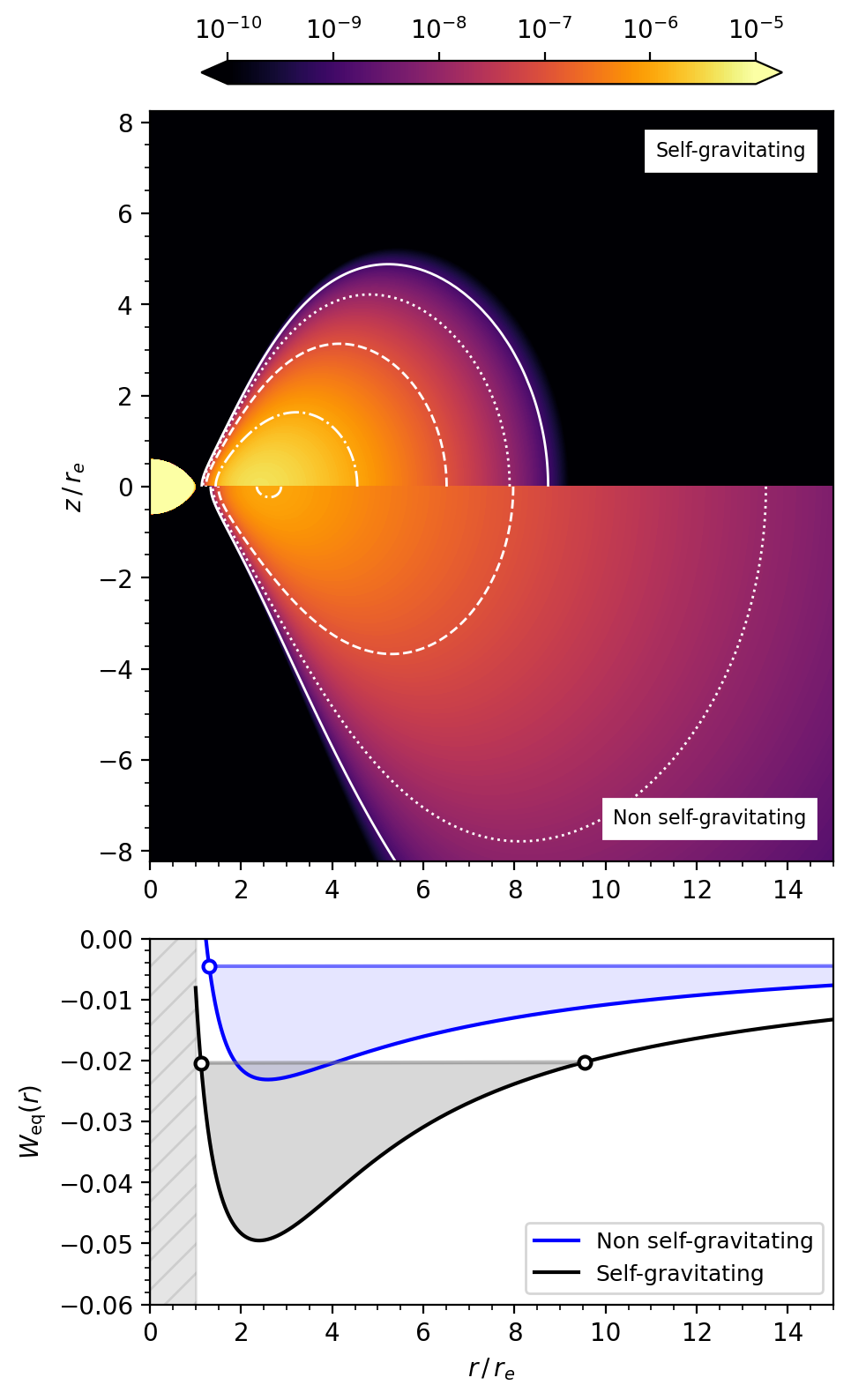}
\caption{Same as Figure~\ref{fig:comparison low mass}, with $\kappa = 7.78$ and
    $M\disk = 0.969M_\odot$. In the upper panel, an additional contour
    level $\rho_0=10^{-6}$ is shown with a white dash-dotted line.}
\label{fig:comparison high mass}
\end{figure}

\begin{figure}
\centering
\includegraphics[width=\linewidth]{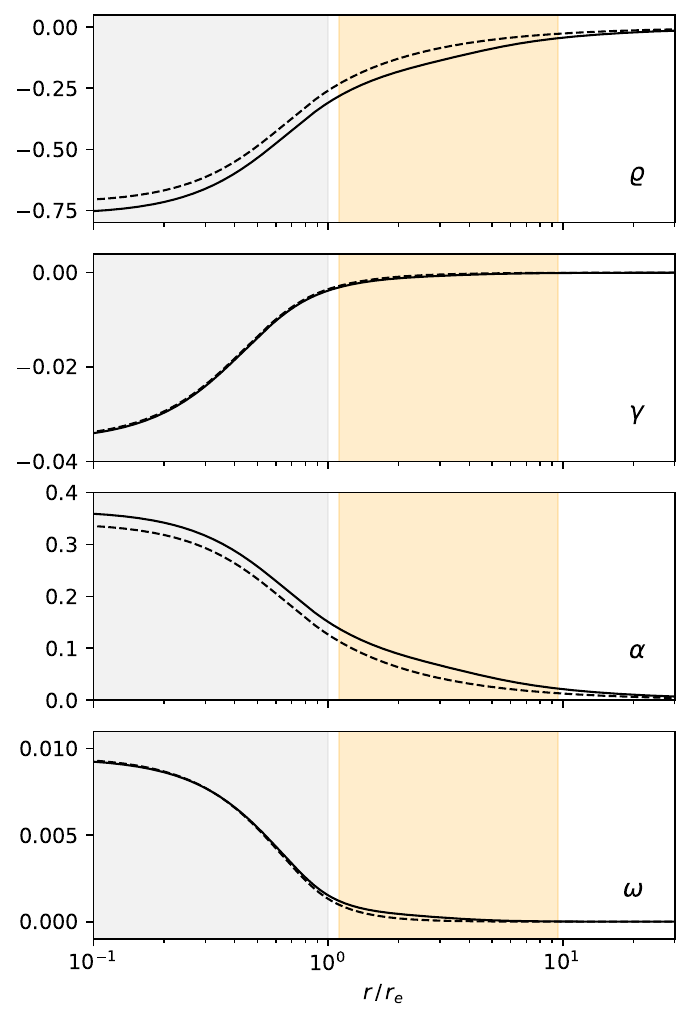}
\caption{Comparison of the metric functions $\varrho$, $\gamma$, $\alpha$, and
    $\omega$ (see Eq.~\eqref{eq:metric} for definition) on the equatorial plane
    with (solid) and without (dashed) the disk self-gravity. Interior of the
    neutron star and the disk are shown with gray and orange shades,
    respectively. Model parameters and the self-gravitating solution is same as
    the one shown in Figure~\ref{fig:comparison high mass}.}
\label{fig:metric comparison}
\end{figure}

\begin{figure}
\centering
\includegraphics[width=\linewidth]{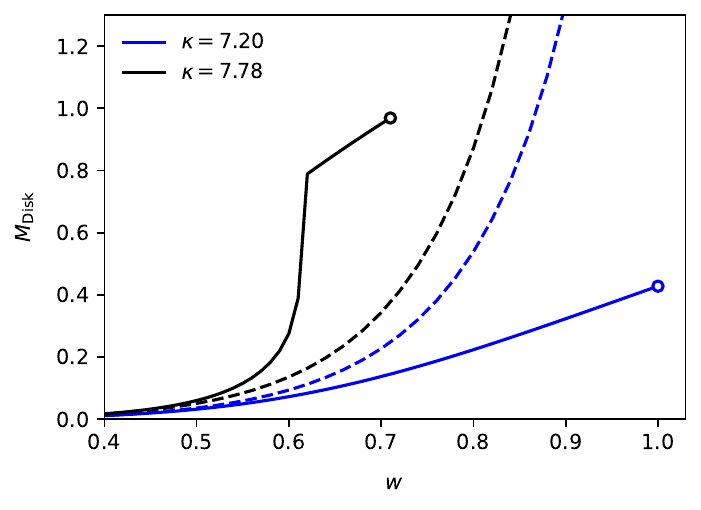}
\caption{Rest mass of the disk (in units of the solar mass) versus the depth
    parameter $w$. Non self-gravitating disk solutions are shown with dashed
    lines while the self-gravitating solutions are shown with solid lines. Model
    parameters are $\rhomax=1.0\e{-3}$, $r_p/r_e = 0.60$, and $a=0.0$. A kink of
    the $\kappa=7.78$ curve near $M\disk=0.8M_\odot$ is a parametrization
    artifact; see discussions in Sec.~\ref{sec:importance of self gravity}.}
\label{fig:mass comparisons}
\end{figure}

So far, we have mainly focused on the influence of a self-gravitating disk on
the neutron star, where the matter distribution of the disk would also
reorganize itself due to its self-contribution to the gravitational field. In
this section, we attempt to assess the impact of the self-gravity of the disk on
itself, compared to the cases in which we ignore it in calculations.

We construct an equilibrium model with $\rhomax = 1.0\e{-3}$, $r_p/r_e = 0.6$,
$\kappa=7.88$, $a=0.0$, and $w=0.54$, yielding the disk rest mass $M\disk=0.101
M_\odot$ where $M\ns=1.66 M_\odot$. Next, using the same set of parameters for
the neutron star, we construct another equilibrium solution with a
non-self-gravitating disk having the same rest mass. Figure~\ref{fig:comparison
low mass} compares these two cases. The self-gravity of the disk drags the outer
edge $r_\text{out}$ inward while the inner edge $r_\text{in}$ does not shift as
much. It can be seen that the spatial distribution of the low-density outer
envelope of the disk is notably compressed toward the disk center in the
self-gravitating case. The difference in the coordinate radius (rest mass) of
the neutron star between these two models is 0.2\% (0.01\%), which is very small
compared to the relative change of the disk structure shown in
Figure~\ref{fig:comparison low mass}. Therefore, we conclude that this spatial
compression of the disk is mostly, if not entirely, due to its self-gravity. The
effective potential at the center of the disk $W_\text{center}$ shows $\sim$
10\% difference in the presence of the disk self-gravity, which agrees in orders
of magnitude with $M\disk/M\ns\approx 0.06$.

We show a more extreme case with $M\disk = 0.97 M_\odot$ ($\kappa = 7.78$) in
Figure~\ref{fig:comparison high mass}. The difference of the neutron star
between the self-gravitating and non-self-gravitating disk solutions in this
case is 3\% in radius and 0.2\% in mass, while the spatial structure of the disk
undergoes a much more dramatic change.
Using the same solution, Figure~\ref{fig:metric comparison} shows the metric
functions $\varrho$, $\gamma$, $\alpha$, $\omega$ as a function of $r$ on the
equatorial plane along with the case without the disk self-gravity. Due to an
increased total gravitational mass of the system, all metric functions have
larger absolute values compared to the non-self-gravitating case. One exception
is the frame dragging potential $\omega$, which shows a slight increase in $r
\leq 0.34r_e$. At the center of the neutron star, $\varrho$ and $\alpha$ show
about 7\% difference, while $\gamma$ and $\omega$ undergo about 8-9 times
smaller changes. In terms of the relative change, both metric functions
$\varrho$ and $\gamma$ show the most deviation at $r = 5.20r_e$, having as about
twice as larger absolute values. The metric function $\alpha$ shows the maximum
70\% increase in its value at $r = 5.32r_e$, where the function $\omega$ shows
its maximum increase of 6.12 times at $r = 6.44r_e$. Note from
Figure~\ref{fig:comparison high mass} that the center of the disk is located at
$r=2.40r_e$, which does not conform with the locations at which metric functions
show the largest deviation from the non-self-gravitating disk model.

In Figure~\ref{fig:mass comparisons}, we plot the rest mass of the disk versus
the disk depth parameter $w$ for a self-gravitating disk and a
non-self-gravitating case, for $\kappa=7.20$ and $\kappa=7.78$ with other model
parameter fixed as $\rhomax=1.0\e{-3}$, $r_p/r_e = 0.60$, $a=0.0$. Since a
non-self-gravitating disk with $w=1$ corresponds to an open (infinite) disk, the
mass of the disk diverges as $w\rightarrow 1$ (dashed lines). The discrepancy
between self-gravitating and non-self-gravitating cases is already clearly
visible from $M\disk \approx 0.1 M_\odot$.
A kink present on the $\kappa=7.78$ (black solid line) plot in
Figure~\ref{fig:mass comparisons} is a parametrization artifact from its
definition Eq.~\eqref{eq:disk depth upper bound}. Due to the self-gravity of the
disk with a large mass, $W_\text{eq}(r_e)$ can drop to a negative value and
parametrization upper bound is changed from a fixed point (zero) to
$W_\text{eq}(r_e)$, inducing a discontinuous jump in the slope $dM\disk/dw$.
We also note that when adopting the parametrization of the disk depth in terms
of the variable $w$, the deviation of the mass of a self-gravitating disk from a
non-self-gravitating case can be both positive or negative; a self-gravitating
solution can lead to either a more massive disk or a less massive disk,
depending on other physical parameters associated with the equilibrium state.

\section{Summary and conclusion}
\label{sec:summary}

Extending the scheme of \cite{Komatsu1989a} with an extra step constructing an
equilibrium disk during iterations, we have constructed classes of numerical
models describing a self-gravitating equilibrium disk orbiting around a rapidly
rotating neutron star. Our formulation makes use of the specific angular
momentum $l$---instead of the coordinate angular velocity $\Omega$---in
prescribing the rotation profile of the disk, facilitating parameter studies on
the disk rotation in a more direct manner.

Since the explorable parameter space of the star-disk system is much wider than
that of an isolated rotating neutron star, the scope of this paper was somewhat
limited, namely assuming a uniformly rotating neutron star and the disk
following the angular momentum distribution given by Eq.~\eqref{eq:disk rotation
law} on the equator. Despite these simplifications, several impacts of the
self-gravity of the disk on the system could have been identified:
\begin{itemize}[leftmargin=1em, labelwidth=1em, itemsep=0.5ex]
\item In an equilibrium state with a given central rest mass density, the
    neutron star at the center decreases slightly in its mass and radius, and
    its rotation becomes slower. These basic findings are consistent with the
    results from \cite{Nishida1992}.
\item Compared to a non-self-gravitating case, the disk is radially compressed
    toward the star. Outer low-density regions are affected most while the
    change is marginal near the inner edge and the center of the disk.
\item Constant angular momentum disk is preferable for the formation of a more
    massive and thick disk near the neutron star. Sub- or nearly-Keplerian disks
    in general show lower rest mass density and more extended geometry, which
    may make it very difficult to accurately compute an equilibrium solution for
    which the self-gravity of the disk is large enough.
\item Roughly speaking, the self-gravity of the disk impacts its internal
    structure with the orders of $M\disk/M\ns$, while little affecting the mass
    or radius of the neutron star.
\end{itemize}

The uniformly rotating assumption on the neutron star greatly simplifies the
problem, but at the same time largely restricting us from resolving
radius-dependent effects by the gravitational pull of the disk to the interior
of the neutron star. A more realistic, differential rotation law
\cite[e.g.][]{Galeazzi2012,Uryu2017,Bauswein:2017aur,Bozzola2018,Iosif2021} can
be used for an improved modeling of the rotating neutron star. Also, we only
have considered a simple power law (Eq.~\eqref{eq:disk rotation law}) as a
representative example of a non-constant angular momentum distribution of the
disk, where several alternate prescriptions are available, such as adopting a
different form of angular momentum distribution \citep{Penna:2013zga,Qian2009}
or specifying a nontrivial functional relationship between the angular velocity
$\Omega$ and the specific angular momentum $l$
\citep{Witzany:2017zrx,Wielgus2015}. Further research with an extended set of
rotation profiles of both the neutron star and the disk will allow us to examine
more diverse configurations of the system.

Several potential improvements for speeding up the computational procedure
(described in Sec.~\ref{sec:method outline}) can be available, such as an
adaptive control of the convergence threshold $\Delta$ and the increment in the
disk depth parameter $w$ when approaching a solution, or applying the successive
over-relaxation method \citep{varga_matrix_2000,hageman_applied_2004} over each
iterations.

On the matter physics, further improvements can be made by using a more
realistic equation of state in the typical rest mass density range of the disk
observed in the current work ($\rho_0 = 10^{6}-10^{11}\si{\gram\per\cubic\cm}$),
as well as adopting a tabulated, nuclear equation of state for the neutron star
rather than using a simple polytrope. The remnant of the binary neutron star
merger or an accreting proto-neutron star is expected to be born hot exceeding 1
MeV $\approx 10^{10}\text{ K}$ \citep{Bernuzzi:2020tgt,prakash1997,Haensel2007}.
A realistic equilibrium model aiming to imitate such hot accreting states needs
to incorporate thermal contributions as well, for which a recently developed
approach by \cite{Camelio:2019rsz,Camelio:2020mdi} for non-barotropic
equilibrium solutions can be useful.

Since our models do not take energy and pressure contributions from magnetic
fields into account, they are only applicable to an unmagnetized or at most
weakly magnetized system. In realistic scenarios, magnetic fields can change the
equilibrium configuration \cite[e.g.][]{Komissarov:2006nz} or would trigger
instabilities so that such systems evolve into accreting states. The dynamical
stability of our results is still under question as well. These aspects can be
further investigated by feeding our model to an evolution code as initial data
and running general relativistic hydrodynamics simulations, which we aim as
future work.

\section*{Acknowledgements}

YK thanks to Chan Park for his assistance and comments, and is grateful to Isaac
Legred and Michael Pajkos for helpful discussions.
JK acknowledges that this research was supported by Basic Science Research
Program through the National Research Foundation of Korea (NRF) funded by the
Ministry of Education (NRF-2021R1I1A2050775). HIK acknowledges that this
research was supported by Basic Science Research Program through the National
Research Foundation of Korea (NRF) funded by the Ministry of Education through
the Center for Quantum Spacetime (CQUeST) of Sogang University
(NRF-2020R1A6A1A03047877). HML was supported by the NRF grant No.
2021M3F7A1082056. We acknowledge the hospitality at APCTP where part of this
work was done. Computations were performed on the gmunu cluster at the Korea
Astronomy and Space Science Institute.
Figures in this article were produced using Matplotlib \citep{matplotlib}, Numpy
\citep{numpy}, and Scipy \citep{scipy} packages.
We also thank the anonymous referee for valuable comments and insightful
suggestions.

\section*{Data Availability}

The computed models which are listed in Table~\ref{tab:model lists} and assigned
with Figures (i.e. all entries except A4, B4, D3) are publicly available on
Zenodo \citep{dataset}. Other data underlying this article will be shared upon
request to the corresponding author.

\bibliographystyle{mnras}
\bibliography{references}

\appendix

\section{Detailed code comparison results}
\label{sec:detailed code comparison}

In Table~\ref{tab:code validation}, we list normalized values of the rest mass
$\bar{M}_0 \equiv K^{-1/2(\Gamma-1)} M_0$ and the total angular momentum
$\bar{J}\equiv K^{-1/(\Gamma-1)}J$ computed with our code for a set of rotating
neutron star models used in \cite{Nozawa1998}. All models were computed with the
grid resolution $(N_s, N_\mu) = (401,201)$. Neutron star parameters $(\rhomax,
r_p/r_e)$ corresponding to each model can be looked up from \cite{Nozawa1998}.

\begin{table}
\centering
\caption{Normalized rest mass $\bar{M}_0$ and total angular momentum $\bar{J}$
for a list of neutron star models. Model names and their parameters follow the
naming notation of \citet{Nozawa1998} so that the results can be compared with
Table 1 and Table 2 therein. For the last four models, we display results with
additional significant figures to compare with Table 8 and Table 9 of the
reference.}
\label{tab:code validation}
\begin{tabular}{ lll }
    \hline
    Model & $\bar{M}_0$ & $\bar{J}$ \\
    \hline\hline
    N05sr & 1.52e-01 & 2.18e-03 \\
    N05rr & 1.59e-01 & 1.07e-02 \\
    N075sr & 1.60e-01 & 2.86e-03 \\
    N075mr & 1.87e-01 & 1.74e-02 \\
    N10sr & 1.82e-01 & 4.21e-03 \\
    N10rr & 1.73e-01 & 9.49e-03 \\
    N15sr & 2.08e-01 & 1.02e-02 \\
    N15rr & 2.77e-01 & 2.13e-02 \\
    \hline
    N15sn & 9.7621e-05 & 3.5979e-07 \\
    N15mr & 3.0421e-01 & 3.8759e-02 \\
    N05sn & 6.2995e-08 & 1.3174e-13 \\ 
    N05mr & 1.8259e-01 & 1.7221e-02\\ 
    \hline
\end{tabular}
\end{table}

\bsp	%
\label{lastpage}
\end{document}